\pdfoutput=1
\documentclass[journal,onecolumn]{IEEEtran}

\newcommand{\descr}[1]{\smallskip \noindent \textbf{#1}}
\newcommand{\descrit}[1]{\smallskip \noindent \textbf{\textit{#1}}}
\usepackage{color}
\usepackage[hyphens]{url}
\usepackage{hyperref}
\usepackage[subtle]{savetrees}
\usepackage{fancyhdr}
\usepackage{bbm}
\usepackage{xspace}
\usepackage{enumerate}
\usepackage{lipsum}
\usepackage{multirow}
\usepackage[shortlabels]{enumitem}
\usepackage{amssymb}
\usepackage{makecell}
\usepackage{booktabs} 
\usepackage{amsmath}
\usepackage{mathbbol}
\usepackage{float}

\usepackage{microtype}
\usepackage[left=0.6in,right=0.6in,top=0.5in,bottom=0.5in,includehead,includefoot]{geometry}
\usepackage{graphicx}
\usepackage{subcaption}
\usepackage{mathtools}
\newcommand{\ceil}[1]{\left\lceil #1 \right\rceil}
\newcommand{\floor}[1]{\left\lfloor #1 \right\rfloor}
\allowdisplaybreaks 
\usepackage{amsthm} 
\usepackage{graphicx}
\usepackage{comment}
\usepackage{xcolor}
\usepackage{alltt}
\usepackage{arydshln}
\usepackage{pifont}

\makeatletter

\usepackage{tikz}
\usetikzlibrary{decorations.pathreplacing,calc}

\usepackage{algorithm,algpseudocode}
\usepackage{setspace}
\usepackage{caption}

\usepackage{subcaption}
\newcommand{\sys}{\textsc{Rhode}\xspace}
\newcommand{\poseidon}{\textsc{Poseidon}\xspace}

\usepackage{tabularx}
\usepackage[T1]{fontenc}
\usepackage[utf8]{inputenc}
\usepackage{bm}
\definecolor{mygreen}{rgb}{0.0, 0.5, 0.0}
\DeclareCaptionType{copyrightbox}

\newif\ifcomment
\commenttrue 

\ifcomment
	\newcommand{\ap}[1]{\textbf{\em\color{blue}[AP: #1]}}

\else  
    \newcommand\ap[1]{}
\fi

\newcommand{\sigmoid}{\ensuremath{\frac{1}{1+e^{-x}}}}

\newcommand{\relu}{\ensuremath{max(0,x)}}
\newcommand{\softplus}{\ensuremath{ln(1+e^x)}}
\newcommand{\tanhh}{\ensuremath{\frac{e^x-e^{-x}}{e^x+e^{-x}}}}

\begin{document}

\title{Privacy-Preserving Federated Recurrent Neural Networks}
\author{
    \IEEEauthorblockN{Sinem Sav\IEEEauthorrefmark{1}, Abdulrahman Diaa\IEEEauthorrefmark{2}, Apostolos Pyrgelis\IEEEauthorrefmark{1}, Jean-Philippe Bossuat\IEEEauthorrefmark{3}, and Jean-Pierre Hubaux\IEEEauthorrefmark{1}\IEEEauthorrefmark{3}\\ }
    \IEEEauthorblockA{\IEEEauthorrefmark{1}École Polytechnique Fédérale de Lausanne (EPFL)\\
    }
     \IEEEauthorblockA{\IEEEauthorrefmark{2}University of Waterloo\\
    }
    \IEEEauthorblockA{\IEEEauthorrefmark{3}Tune Insight SA\\
    }
}

\maketitle
\thispagestyle{plain}
\pagestyle{plain}

\begin{abstract}

We present \sys, a novel system that enables privacy-preserving training of and prediction on Recurrent Neural Networks (RNNs) in a cross-silo federated learning setting by relying on multiparty homomorphic encryption. \sys preserves the confidentiality of the training data, the model, and the prediction data; and it mitigates federated learning attacks that target the gradients under a passive-adversary threat model. We propose a packing scheme, \textit{multi-dimensional packing}, for a better utilization of Single Instruction, Multiple Data (SIMD) operations under encryption. With multi-dimensional packing, \sys enables the efficient processing, in parallel, of a batch of samples. To avoid the exploding gradients problem, \sys provides several clipping approximations for performing gradient clipping under encryption. We experimentally show that the model performance with \sys remains similar to non-secure solutions both for homogeneous and heterogeneous data distribution among the data holders. Our experimental evaluation shows that \sys scales linearly with the number of data holders and the number of timesteps, sub-linearly and sub-quadratically with the number of features and the number of hidden units of RNNs, respectively. To the best of our knowledge, \sys is the first system that provides the building blocks for the training of RNNs and its variants, under encryption in a federated learning setting.
\end{abstract}

\section{Introduction}\label{sec:intro}

Collecting and mining sequential data, e.g., time-series, has become the base for numerous real-life applications in the domains of healthcare, energy, and finance. For instance, in personalized healthcare, electrocardiograms (ECGs) are used to monitor patients and to detect heartbeat arrhythmia; and, in finance, historical economic records are mined to forecast stock prices. The state-of-the-art time-series mining is achieved by employing machine learning (ML) algorithms that extract useful patterns from the data. A well-known class of algorithms widely used for learning on time-series data is recurrent neural networks (RNNs)~\cite{Rumelhart1986,Elman1990} and its variants, e.g., Gated Recurrent Units (GRUs) and Long-short Term Memory (LSTMs), that are capable of modelling high-dimensional and non-linear relationships in dynamic systems~\cite{Schafer2006}.


Training effective RNNs for time-series tasks requires large amounts of data that are usually generated by multiple sources and scattered across several data holders. Sharing or centralizing this data is difficult because it is privacy sensitive~\cite{Zhang_BigDataSecurity}. For example, smart-meter data might leak householders' identities~\cite{Buchmann2013} and their activities~\cite{Markham2010}, location data reveal information about peoples' lifestyles and beliefs~\cite{Shokri2011,Shokri20112,Olteanu2017,Hui2011}, and healthcare time-series are by nature private information~\cite{Barni2011,Liu2020}. Additionally, privacy regulations enforce strict rules for sharing this type of data~\cite{HIPAA,GDPR}. As a result, privacy-preserving data sharing and collaborative ML solutions have gained a significant importance~\cite{Liu2020,Boulemtafes,Hong2013}.

Federated learning (FL) is a machine learning paradigm that enables privacy-preserving collaborative training of ML models. With FL, data holders retain their data in their own premises and share model updates with an aggregation server that coordinates the learning over their joint data~\cite{federatedLearning1,Konecny2016,Konency2016fed}. FL has been applied in many time-series applications by using RNNs to forecast the energy load~\cite{Fekri2022}, to predict the next word typed on mobile phones~\cite{Hard2018,Chen2019}, and to classify cancer in healthcare systems~\cite{Gandhi2021}. Although FL provides collaborative learning without data sharing, it still raises privacy concerns: The shared intermediate model is vulnerable to privacy attacks that can reconstruct parties' input data or infer the membership of data samples in the training set, hence needs to be protected during the training process~\cite{Hitaj2017,Wang2019,Zhu2019,Melis2019,Nasr2019,Zhao2020,Jonas2020,Jin2021,Wainakh2022,Deng2021,Dimitrov2022}. Moreover, it was recently shown that RNNs are particularly vulnerable to inference attacks (e.g., membership inference) compared to traditional neural networks~\cite{yang2021privacy}.

To overcome these shortcomings, several works propose integrating protection mechanisms such as differential privacy (DP), secure multiparty computation (SMC), or homomorphic encryption (HE) in the FL process. Each of these techniques, however, introduces a different trade-off: DP solutions decrease the utility of the ML model --- in particular for RNNS~\cite{yang2021privacy} --- while their level of practical privacy protection remains unclear~\cite{jayaraman2018distributed}; SMC solutions lack scalability in terms of circuit complexity and number of parties due to high communication overheads~\cite{Goyal2022,keller2022secure}; and HE solutions do not scale with the model complexity due to high computational overhead. Recently, a multifaceted approach, termed multiparty homomorphic encryption (MHE)~\cite{Franklin1996,Asharov2011}, where the private key is secret-shared among the parties, has been used by several works to better balance the aforementioned trade-offs with privacy-preserving federated learning~\cite{spindle,poseidon}.
However, these solutions focus solely on the training of generalized linear models or feed-forward neural networks and do not tackle the training of RNNs, which introduces its own additional challenges: (i) the sequential computations make the training procedure slow and hard to parallelize, and (ii) the long-term dependencies lead to exploding/vanishing gradients~\cite{Pascanu2013,Bengio1994}.

By building upon MHE-based works~\cite{spindle,poseidon}, we design \sys, a novel system that enables the training of and predictions on RNNs and its variants in a cross-silo federated learning setting with $N$ data holders \emph{by performing all the computations under homomorphic encryption}. \sys protects the confidentiality of the training data, the local gradients, and the global model under a passive adversary model (a common assumption for cross-silo settings) that allows collusions of up to $N-1$ data holders. To achieve this, both the local gradients and the FL model remain encrypted throughout \sys's pipeline. Hence, \sys mitigates passive FL attacks that target the gradients or the global model by concealing all operations throughout the learning. To accelerate RNN training, \sys enables mini-batch training by relying on a packing scheme that reduces the processing time of a batch. To address the problem of exploding gradients, \sys uses efficient polynomial approximations of the gradient clipping function. After training, \sys enables predictions-as-a-service (PaaS) to a querier that obliviously evaluates the trained model on its private inputs and obtains the prediction results. We implement \sys and our experimental evaluation shows that its accuracy is on par with non-private centralized and decentralized solutions. Moreover, it scales sub-linearly with the number of features, sub-quadratically with the hidden RNN dimension, and linearly with the timesteps, the batch size and the number of data holders. For example, \sys can process $256$K samples distributed among 10 data holders with an RNN of $32$ hidden units and $4$ timesteps, over $100$ global training iterations, in $\sim 1.5$ hours. In summary, our contributions are the following:
\begin{itemize}
    \item We present \sys, a novel system for the training of RNNs in a cross-silo FL setting that conceals all intermediate values and the model that are communicated in the FL process by relying on MHE. After training, \sys enables oblivious predictions to a querier that provides its encrypted inputs.
    \item We design a multi-dimensional packing scheme that enables efficient encrypted matrix operations suitable for mini-batch training. We show that matrix multiplication with our scheme improves the state-of-the-art~\cite{Jiang2018,poseidon} in terms of throughput proportionally to the dimension of the matrices. Moreover, our packing scheme preserves the number of the cryptographic keys regardless of the matrix size.
    \item We propose and evaluate polynomial approximations for performing gradient clipping under encryption to alleviate the problem of exploding gradients that is inherent to RNNs. 
    \item We experimentally evaluate \sys and show that:
    \begin{itemize}[leftmargin=*]
        \item[-] It scales sub-linearly with the number of features, sub-quadratically with the hidden RNN dimension, linearly with the timesteps or the batch size, and sub-linearly (or linearly) with the number of data holders depending on the number of samples in the dataset.
        \item[-]Its homomorphic operations and building blocks have negligible effect on the model performance for common time-series forecasting benchmarks, with both homogeneous and heterogeneous (imbalanced) data distribution among the parties.
    \end{itemize}
   
\end{itemize}

\noindent To the best of our knowledge, \sys is the first system that enables the training of and prediction on RNNs in a cross-silo FL setting under encryption. \sys mitigates passive FL attacks during training and preserves the privacy of the data holders’ data, the intermediate model, the querier's evaluation data, and the final model. 
\section{Related Work}\label{sec:related}
We review the related work on privacy-preserving time-series, as well as on privacy-preserving federated training of, and prediction on machine learning (ML) models.
\subsection{Privacy-Preserving Time-Series} 

Previous works aiming to protect the privacy of time-series data employed secure aggregation techniques to enable the computation of simple statistics and analytics~\cite{Kursawe2011,Melis2016,Pyrgelis2016}, whereas others combined secure aggregation with differential privacy to bound the leakage stemming from the computations~\cite{Shi2011,Brown2013}. More recently, Liu et al.~\cite{Liu2019} employed secure multi-party computation (SMC) techniques for privacy-preserving collaborative medical time-series analysis.
Dauterman et al.~\cite{dautermann2022} use function secret sharing (FSS) to support various functionalities, e.g., multi-predicate filtering.
All these authors focus on simple collaborative time-series tasks and not on the privacy-preserving training of ML models on time-series data, which is the main focus of this work.


\subsection{Privacy-Preserving Federated Training of Machine Learning Models}
Federated learning (FL) enables different data holders to perform the collective training of a ML model while maintaining the data in their local premises and communicating only the local model updates to an aggregation server~\cite{federatedLearning1,Konecny2016,Konency2016fed}. FL was applied on many time-series applications such as load forecasting~\cite{Fekri2022}, natural language processing~\cite{Chen2019,Hard2018}, traffic flow prediction,~\cite{Liu2020Traffic}, and healthcare systems~\cite{Gandhi2021}. However, recent research has shown that sharing gradients or local model updates in FL leads to severe privacy leakage because membership inference and data/label reconstruction attacks are effective~\cite{Hitaj2017,Wang2019,Zhu2019,Melis2019,Nasr2019,Zhao2020,Jonas2020,Jin2021,Wainakh2022,Deng2021,Dimitrov2022}.

One common approach for privacy-preserving \emph{decentralized} or \emph{federated} ML is to use secure aggregation protocols that hide the data holders' model updates while only revealing their combination to the server. Various such protocols have been proposed based on HE~\cite{Phong2017,Phong2018,liu2019secure}, functional encryption~\cite{xu2019hybridalpha,9860595}, or SMC~\cite{bonawitz2017practical, bell2020secure}. However, the aggregated global model remains vulnerable to privacy attacks such as model inversion~\cite{fredrikson2015model} or membership inference~\cite{shokri2017membership}. Other works integrate differential privacy (DP) into the learning phase by injecting noise on the intermediate model updates~\cite{shokri2015privacy,Nvidia_Fed,abadi2016deep,McMahan2018,Wei2020,wu2019value} or by combining it with secure aggregation techniques~\cite{jayaraman2018distributed,truex2019hybrid,stevens2022efficient} such that the aggregated global model meets the DP requirements. McMahan et al. recently applied DP to RNNs to protect user-level privacy~\cite{mcmahan2018LSTM}, and Wen et al. applied local DP to a FL framework for energy-theft detection~\cite{Wen2021}. However, DP introduces a privacy vs. utility trade-off that is hard to parameterize. Yang et al. show that perturbing the gradients during training or the final model weights to achieve DP guarantees results in higher utility loss for RNNs than other neural networks~\cite{yang2021privacy}.


Another line of research applies SMC to protect the privacy of the \textit{full} decentralized ML pipeline~\cite{nikolaenko2013privacy,gascon2017privacy,giacomelli2018privacy,Akavia_WAHC,schoppmann2019make,bogdanov2016rmind,Cho_GWAS,SecureML,quotient,mohassel2018aby,wagh2019securenn,falcon,Chen,flash,trident,Knott2021,Tan2021CryptGPUFP}. However, SMC solutions require data holders to secret-share their data among a limited number of computing parties and their confidentiality is ensured under non-collusion assumptions (typically, honest majority~\cite{SecureML,wagh2019securenn,quotient,flash,trident,Chen}) among them. Furthermore, they introduce high communication overhead and scale poorly in terms of the circuit complexity or the number of parties~\cite{Goyal2022,keller2022secure}. Overall, existing SMC solutions enable the training of regression models and feed-forward neural networks, whereas RNNs have not been studied. One recent work, which relies on additive secret-sharing, investigated the training of recurrent neural networks as a case study~\cite{Zhang2020}, but without evaluating runtime performance or scalability.

Some works employ homomorphic encryption (HE) to encrypt the clients' data, before outsourcing it to a server that performs the model training~\cite{Zhang2018}. Zheng et al. use SMC in combination with HE for maliciously secure FL~\cite{zheng2019helen,Zheng2021}. None of these works, however, address the learning of sequential patterns over the data. To balance the trade-off between privacy, utility, and performance, several works apply multiparty homomorphic encryption (MHE)~\cite{Asharov2011} to realize decentralized trust and to enable efficient collaborative homomorphic operations, such as collective bootstrapping~\cite{mouchet2019distributedbfv}. These works apply MHE to FL to conceal any information exchanged during the learning process~\cite{spindle,poseidon}. However, their focus is on generalized linear models~\cite{spindle} or feed-forward neural networks~\cite{poseidon}, and not the training of RNNs. For instance, the packing scheme and the matrix/vector operations employed in~\cite{poseidon} are tailored for the processing of a single sample (thus eliminating explicit transpose and matrix multiplications; see Section~\ref{sec:microbenchmarks} for comparisons demonstrating the unsuitability of this approach for training RNNs). Whereas in this work, we propose a novel packing scheme suitable for mini-batch RNN training and enable matrix transpose and matrix multiplications. Moreover, we propose several polynomial approximations to enable critical operations for RNNs, such as gradient clipping, under MHE.

A different line of research is focused on the private aggregation of teacher ensembles (PATE)~\cite{papernot2017semisupervised,Papernot2018}, and its variants~\cite{long2021gpate,yoon2018pategan}; focusing mainly on knowledge transfer by combining DP and generative adversarial networks (GANs). 
Our work diverges from these works due to their assumption of publicly available datasets and their designs that target knowledge transfer. Choquette-Choo et al. recently proposed CaPC Learning that enables confidential and private collaborative learning by relying on SMC, HE, and DP, in order to increase the utility of each data holder's local ML model~\cite{Choquette2021}. 
However, CaPC Learning is evaluated only on feed-forward neural networks and the number of parties needs to be large enough to obtain better utility/privacy guarantees. Our work decouples privacy from the amount of data or the number of parties.

Overall, \sys enables the training of RNNs in a cross-silo FL while protecting the confidentiality of the training data, the local gradients, and the global model in a passive threat model with up to $N-1$ collusions among the data holders. By design, \sys mitigates passive FL inference attacks during training, as all intermediate values that are communicated remain encrypted and as it enables all the computations of the FL pipeline \textit{under encryption}.

\subsection{Privacy-Preserving Prediction on Machine Learning Models}
Privacy-preserving prediction on ML models plays an important role for Prediction-as-a-Service (Paas) solutions whose aim is to protect the model from the client (who queries the model) and the client's evaluation data from the service provider. A series of works employ SMC or HE (or a combination of them) to enable privacy of both the data and the model holder~\cite{CryptoNets,MiniONN,Gazelle,blaze,riazi2019xonn,mishra2020,cheetah2022,boemer2019ngraph,cryptoDL,Deevashwer2020}. These works, however, develop and improve cryptographic primitives that support the evaluation of feed-forward neural networks. Only a few recent studies focus on privacy-preserving prediction on RNNs by relying on several cryptographic primitives such as HE~\cite{Maya2020,10.1145/3488932.3523253}, two-party computation (2PC)~\cite{Rathee2021SiRnn}, secure multiparty computation~\cite{Feng2020}, and hybrid approaches combining HE and garbled circuits~\cite{feng2021cryptogru}. 
Similar to these works, \sys protects the confidentiality of the model and the client's evaluation data during the prediction phase. However, our work is broader, as it also enables federated \textit{training} under encryption.

\section{Background}\label{sec:background}
We present preliminaries on multiparty homomorphic encryption (MHE), federated learning, and recurrent neural network (RNNs).
\subsection{Multiparty Homomorphic Encryption}\label{sec:mhe}

In this work, we rely on multiparty homomorphic encryption (MHE) to enable distributed cryptographic operations. We employ the multiparty variant of Cheon-Kim-Kim-Song (CKKS), a leveled homomorphic-encryption scheme based on the \textit{ring learning with errors} (RLWE) problem~\cite{cheon2017homomorphic}, that is proposed by Mouchet et al.~\cite{mouchet2019distributedbfv}. This scheme is secure against post-quantum attacks, enables floating-point arithmetic, and tolerates collusions of up to $N-1$ parties, under a passive threat model. Below, we summarize the most important operations of this scheme:

\descr{Key Generation:} Each data holder $i$ locally generates a \textit{secret} key ($sk_i$). The \textit{public} key ($pk$) and evaluation keys ($[ek]$) are collectively generated from the data holders' secret keys ($\textsf{CKeyGen}([sk_i])$). $[ek]$ is a set of special public keys that are required for the execution of homomorphic multiplications and rotations. With this scheme, any data holder can encrypt and locally perform homomorphic operations by using $pk$ and $[ek]$, but decrypting a ciphertext requires the collaboration among all data holders ($\textsf{CDecrypt}(\mathbf{c}_{pk},[sk_i])$). We denote any ciphertext encrypted under $pk$ with bold-face $\mathbf{c}_{pk}$ throughout this paper. When there is no ambiguity, we omit the sub-index and use $\mathbf{c}$.
    
\descr{Arithmetic Operations and Parallelization:} The CKKS homomorphic encryption scheme enables approximate arithmetic over a vector of complex numbers ($\mathbb{C}^{\mathcal{N}/2}$) for a polynomial ring of dimension $\mathcal{N}$. Encrypted operations over $\mathbb{C}^{\mathcal{N}/2}$ are carried out in a Single Instruction Multiple Data (SIMD) manner, which enables parallelization over the $\mathcal{N}/2$ plaintext vector slots.
    
\descr{Arithmetic Circuit Depth and Collective Bootstrapping:} A fresh ciphertext has an initial level of $\mathcal{L}$ and at most $\mathcal{L}$ multiplications (i.e., an $\mathcal{L}$-depth circuit) can be evaluated on it. When the levels are exhausted, the ciphertext is refreshed with a collective bootstrapping operation ($\textsf{CBootstrap}(\mathbf{c}_{pk},[sk_i])$) that enables further operations. We note here that $\textsf{CBootstrap}(\cdot)$ is a costly operation with approximately $2$ orders of magnitude higher overhead than a homomorphic addition/multiplication.
    
\descr{Slot Rotations:} The slots of a ciphertext can be re-arranged via rotations to the left/right. $\textsf{RotL/R}(\mathbf{c}_{pk}, k)$ homomorphically rotates $\mathbf{c}_{pk}$ to the left/right by $k$ positions using $[ek]$.
    
    
\descr{Collective Key-Switch ($\textsf{CKeySwitch}(\mathbf{c}_{pk}, pk', [sk_i])$):} This operation changes the encryption key of a ciphertext $\mathbf{c}$ from $pk$ to $pk'$. As such, only the holder of the secret key corresponding to $pk'$ can decrypt the result $\mathbf{c}_{pk'}$.
    

\noindent Note that the aforementioned operations are MHE (i.e., interactive) ones if they are termed as `collective' (or if the operation name starts with $\textsf{C}$, e.g., $\textsf{CKeyGen}(\cdot)$), otherwise they are pure CKKS functionalities. For further details about the CKKS scheme, e.g., relinearization and rescaling, we refer the reader to the original work of Cheon et al.~\cite{cheon2017homomorphic}, and for more information about the collective MHE operations to the work of Mouchet et al.~\cite{mouchet2019distributedbfv}.

\subsection{Federated Learning}
Federated Learning (FL) is an emerging collaborative ML approach that enables multiple data holders to train a model without sharing their local data~\cite{federatedLearning1,Konecny2016,Konency2016fed}. With FL, each data holder performs several training iterations on its local data, and the resulting local models are aggregated into a global model via an aggregation server. We summarize the most popular federated learning algorithm, i.e., $\textsf{FedAvg}$, proposed by McMahan et al.~\cite{federatedLearning1} in Algorithm~\ref{algo:fedavg}. ${W}_{i}^{k}$ denotes the model weights of the $i^{th}$ data holder during the $k^{th}$ iteration. If no sub-index is used, then it simply denotes the global model weights at a certain iteration. The server initializes the model weights, sends them to a random subset of data holders $S$ that execute the local gradient descent algorithm ($\textsf{Local Gradient Descent}(\cdot)$; see Section~\ref{sec:solutionOverview} for the details of the algorithm when it is tailored for RNN execution under encryption with \sys) to update their local model. The model updates from each data holder are aggregated to the global model via weighted averaging ($n_i$ denotes the number of local samples at the $i^{th}$ data holder and $n$ the total number of samples). The number of global and local iterations are denoted as $e$ and $l$, respectively. Note that $\textsf{FedAvg}$ is a generalization of the standard $\textsf{FedSGD}$ algorithm with several local iterations performed on each data holder's side.

\begin{algorithm}[h]
\small
\caption{Federated Averaging Algorithm ($\textsf{FedAvg}$)}
\label{algo:fedavg}
\begin{algorithmic}[1]
\State Initialize the global model ${W}^0$ \Comment{Server executes}
\For{$k=0 \rightarrow e-1$} \Comment{Global iterations}

\State Choose $S$ random subset of $N$ data holders
\State  Send ${W}^k$ to each data holder in $S$

 \For{$\ell=0 \rightarrow l-1$} \Comment{Local iterations}
 \State  ${W}_{i}^{k+1} \leftarrow \textsf{Local Gradient Descent}(\cdot)$ \Comment{at each $S_i \in S$}
 \EndFor
\State ${W}^{k+1} \leftarrow \sum_{i=1}^{N} \frac{n_i}{n} { W}_{i}^{k+1}$ \Comment{Server executes}
\EndFor
\end{algorithmic}
\end{algorithm}

\subsection{Recurrent Neural Networks}
Recurrent neural networks (RNNs) are a special type of models that enable learning on sequential data (e.g., time-series)~\cite{Rumelhart1986,Elman1990}. A typical RNN is composed of an input layer that obtains the sequential input to the network, a hidden layer, and an output layer. Contrary to feed-forward neural networks where the information flows from the input to the output layer, in an RNN, the information is fed back to the network through connections between its nodes. The RNN input is a sequence where each item in the sequence is called a timestep. The hidden layer consists of several hidden units (neurons). Each hidden unit has a hidden state that retains the information from the previous timestep. Thus, RNNs capture sequential patterns in the data by processing each timestep sequentially and updating the hidden state. In this way, RNNs use earlier information in the sequence to learn the current output and retain a form of \textit{memory}. A typical RNN, e.g., an Elman network~\cite{Elman1990}, takes a sequential input ($X = x_1,x_2,\cdots,x_T$) of $T$ timesteps and outputs the prediction $y_t$ through the hidden units and an activation function $\varphi(\cdot)$ by sequentially computing the hidden state at time $t$ as:
\begin{equation}
\begin{split}
h_t &= \varphi( h_{t-1}\times {W}  + x_t \times {U} + {b_h})\\
 y_t &= h_t\times{V}  + {b_y}
 \end{split}
 \label{eq:elman}
\end{equation}
where $U$, $W$, and $V$ denote the input-to-hidden, hidden-to-hidden, and hidden-to-output weight matrices, respectively, and $b_h$ and $b_y$ the hidden and output biases. 
Note that the input at each timestep $x_t$ can be itself a $d$-dimensional \textit{feature} vector (or a matrix $b \times d$ for a batch of size $b$). But to avoid notation overflow, we assume that each timestep is a scalar value throughout the paper, unless otherwise stated. The dimensions of $U$, $W$, and $V$ are $d \times h$, $h\times h$ and $h\times o$, where $d$, $h$, and $o$ are the input, hidden, and the output dimensions, respectively. Jordan networks~\cite{JORDAN1997471} are similar to Elman ones but with a slightly different hidden unit, i.e., $h_t = \varphi( y_{t-1}\times {W}  + x_t \times {U} + {b_h})$. Consequently, the size of the weight matrix $W$ in Jordan networks is $o \times h$.

The training of RNNs is an iterative process with a series of forward and backward passes (FP and BP, respectively). FP comprises the prediction by using the formulas in Eq.~\ref{eq:elman}. However, BP is computed via the back-propagation through time (BPTT) algorithm that takes into account the error propagated through the hidden units and computes the gradients by following the chain rule~\cite{Rumelhart1986}. We describe in detail the operations of this process under \sys in Section~\ref{sec:solutionOverview}. Finally, we note that there are various RNN structures, e.g., many-to-one or many-to-many, depending on the input and output layer size (e.g., a many-to-many structure yields outputs for $\kappa$ timesteps). Configuring \sys to support these is trivial and requires simple alterations in its implementation. However, RNN variants in terms of architecture, such as Gated Recurrent Unit (GRU) or Long Short-Term Memory (LSTM), require modifications to the hidden units. Throughout this paper, we consider the traditional RNN architecture that is called simple, vanilla, or an Elman RNN~\cite{Elman1990}, yet our solution can be extended to more complex RNNs architectures (see Section~\ref{sec:complexRNN}).


\section{\sys Design}\label{sec:design}
We introduce the problem and the overview of our solution. We also present the challenges associated with RNN training.

\subsection{Problem Statement}\label{sec:problemStatement}
We first discuss the system and threat model of \sys, as well as its training and prediction objectives. 

\descr{System Model.} We consider a cross-silo FL setting with a moderate number of data holders ($N$) (typically in the range of 2 to 200~\cite{kairouz2019}). The data holders are interconnected in a tree-structured topology for efficient communication (although our system can be adapted to any network topology, e.g., standard FL topology, see Appendix~\ref{sec:extensions}). Each data holder $i$ has its time-series input data $X^i$ and the corresponding output $Y^i$ ($Y^i\in \{0,1,\cdots,n\}$ for n-class classification or $Y^i\in \mathbb{R}$ for regression tasks) for training a joint ML model (horizontal FL). After training, the data holders enable PaaS for a querier $q$ with evaluation data $X^q$. The querier can be one of the data holders or an external entity and obtains $Y^q$, the prediction result. We assume that all data holders are available during execution; dropouts are not supported by \sys due to the underlying cryptographic operations that require all data holders to participate (see Appendix~\ref{sec:extensions} for an extension to tolerate dropouts). 

\descr{Threat Model.} We assume a passive-adversary threat model (which is appropriate and a common threat model for cross-silo settings) that allows collusions between up to $N-1$ parties, including the data holders and the querier. As such, the data holders follow the protocol, but up to $N-1$ of them can share their observations or their inputs to extract information about the honest data holder's inputs. \sys aims at preserving the confidentiality of both the input data and the model during the training and prediction pipelines. We note that our focus is on mitigating passive FL inference attacks that target the model during training. Active adversaries during training or inference attacks targeting the outputs during prediction, e.g., model stealing~\cite{tramer2016stealing}, are out of the scope of this work. We discuss complementary mechanisms that can be used orthogonally to \sys to mitigate such attacks in Appendix~\ref{sec:extensions}.

\descr{Training.} The data holders aim at collaboratively training an RNN model \textit{without} relying on a trusted party and \textit{without} revealing their data to any other party. \sys's objective is to enable this and to mitigate passive FL attacks that target the model or local gradients. As we optimize the communication via a tree-structured topology, the root plays the role of the aggregation server in traditional FL, and for clarity, throughout the paper, we refer to this node as the \textit{aggregation server}.

\descr{Prediction.} \sys's objective is to protect the confidentiality of both the querier's data and the trained RNN model, i.e., the data holders should not learn anything about the querier's evaluation data and the querier should not learn anything about the model other than what it can learn from the prediction output.

\subsection{Solution Overview}\label{sec:solutionOverview}
To fulfill the training and prediction objectives under the given threat model, \sys relies on FL and MHE. To protect data confidentiality and to mitigate the passive FL inference attacks during training, \sys retains the model and all the intermediate values communicated through the network in encrypted form. This implies that all the local RNN operations performed by the data holders need to be carried out under encryption. To efficiently enable this, \sys uses a packing strategy and several cryptographic optimizations that are detailed in Section~\ref{sec:crypto}. To refresh the noise accumulated by the homomorphic operations, \sys performs bootstrapping ($\textsf{CBootstrap}(\cdot)$) when required (depending on the network and security parameters). For clarity, we refer to Algorithm~\ref{algo:fedavg} for our high-level protocol. \sys enables, under encryption, both $\textsf{FedAvg}$ and $\textsf{FedSGD}$ with every data holder included in each iteration (i.e., $|S|=N$) since we consider a cross-silo setting and given that all data holders are available to participate in cryptographic operations. By enabling RNN training, \sys can handle sequential data yet, throughout the paper, we use time-series data as an example.

\subsubsection{Private Training}
The training protocol of \sys operates in four phases: \textbf{Setup}, \textbf{Local Computation}, \textbf{Aggregation}, and \textbf{Model-Update}. We detail these phases hereunder:

\descr{Setup Phase:} In this phase, the necessary cryptographic and learning parameters are decided by the data holders, i.e., the minimum security guarantees, the RNN architecture, and the learning parameters, e.g., the learning rate ($\eta$), the batch size ($b$), and the number of global ($e$) and local iterations ($l$). Then, the data holders generate their secret keys ($sk_i$, for the $i^{th}$ data holder) and collectively generate the corresponding public key ($pk$) with the forenamed MHE scheme (Section~\ref{sec:mhe}). For the training execution to start, the aggregation server initializes the global model (Line 1, Algorithm~\ref{algo:fedavg}), encrypts it with $pk$, and broadcasts it to all data holders.

\descr{Local Computation Phase:} Each data holder receives the encrypted global model weights and begins the \textsf{Local Gradient Descent}($\cdot)$, Line 6, Algorithm~\ref{algo:fedavg}, i.e., the execution of the forward pass (FP) and backward pass (BP) on its local data as detailed in Algorithm~\ref{algo:rnn} for a simple many-to-many RNN with $\kappa=T$ outputs. Note that for different RNN architectures such as GRU or LSTMs, the local gradient descent algorithm should be altered (see Section~\ref{sec:complexRNN}). We denote the encryption of any value, vector, or matrix with bold-face, e.g., $\mathbf{W}$, and for clarity, we omit the indices of the global iteration and the data holder in Algorithm~\ref{algo:fedavg}. $x_t$ is the input $d$-dimensional feature-vector for a timestep $t$ (or a matrix of size $b \times d$ for a batch of size $b$, in which case the outer-vector products in the backward pass become matrix multiplications) and $y_t$ the corresponding output(s). 
$\varphi(\cdot)$ ($\varphi'(\cdot)$) indicates any type of activation function (and its derivative), e.g., $\textsf{Tanh}$. We denote the transpose of a matrix $A$ as $A^T$, the element-wise multiplication with $\odot$, the matrix or vector multiplication with $\times$, and the outer product with $\otimes$. Note that the characteristics of RNNs incur several challenges when training them under encryption. We discuss these challenges in Section~\ref{sec:challenges} and propose several techniques to alleviate them in Section~\ref{sec:crypto}.

\descr{Aggregate Phase:} Data holders collectively aggregate their locally computed models (or gradients). That is, each data holder sends its (encrypted) local model to its parent, and each parent homomorphically sums the received child-models with its own. As such, the aggregation server receives the sum of all the local models.

\descr{Model-Update Phase:} The aggregation server updates the global model by averaging the local models of the data holders (Line 8, Algorithm~\ref{algo:fedavg}). At this phase, the aggregation server also applies gradient clipping (see Section~\ref{sec:approximations} for details). Since the batch size is public information (agreed upon in the Setup Phase), the division for averaging over the encrypted local models becomes a multiplication with a plaintext value. The global model is then broadcast to all data holders for the next Local Computation Phase. 

\noindent Overall, the model training composes the Local Computation, Aggregate, and the Model-Update Phases, which are repeated for a predefined number of global iterations ($e$).

\begin{algorithm}[t]
\small
  \caption{Local Computation Phase (RNN)}
  \label{algo:rnn}

  \begin{algorithmic}[1]
   \Require $\mathbf{U}, \mathbf{W}, \mathbf{V}, \mathbf{h}_{prev}, X = (x_1, x_2, \cdots, x_t), Y = (y_1, y_2, \cdots, y_t)$ 
\noindent \Ensure $\nabla\mathbf{U}, \nabla\mathbf{W}, \nabla\mathbf{V}$
  \State $\mathbf{h}_0 \gets \mathbf{h}_{prev}$
    \For{$t \gets 1 : T$} \Comment{Forward Pass (FP)}
        \State $\mathbf{z}_t \gets \mathbf{h}_{t-1}\times \mathbf{W} +  x_t \times \mathbf{U} $
        \State $\mathbf{h}_t \gets \varphi(\mathbf{z}_t)$ 
        \State $\mathbf{p}_t \gets \mathbf{h}_t \times \mathbf{V} $ 
    \EndFor
    \State $\mathbf{h}_{prev} \gets \mathbf{h}_t$
    \State $\mathbf{dh}_{next} = \mathbf{0}$
    \State $\nabla\mathbf{U}, \nabla\mathbf{W}, \nabla\mathbf{V}  = \mathbf{0}$
    \For{$t \gets T : 1$} \Comment{Backward Pass (BP)}
        \State $\mathbf{dy} \gets \mathbf{p}_t - y_t$ 
        \State $\mathbf{dh} \gets \mathbf{dy} \times \mathbf{V}^T  + \mathbf{dh}_{next}$ 
        
        \State $\mathbf{dz} \gets \mathbf{dh} \odot \varphi'(\mathbf{z}_t) $ 
        
        \State $\mathbf{dh}_{next} \leftarrow \mathbf{dz} \times \mathbf{W}^T$ 
        
        
        \State $\nabla\mathbf{V} \leftarrow \nabla\mathbf{V} +  \mathbf{h}_t \otimes \mathbf{dy} $ 
        
        \State $\nabla\mathbf{U} \leftarrow \nabla\mathbf{U} +  x_t \otimes \mathbf{dz}$ 
        \State $\nabla\mathbf{W} \leftarrow \nabla\mathbf{W} +  \mathbf{h}_{t-1} \otimes \mathbf{dz}$ 
        

    \EndFor
\end{algorithmic}
\end{algorithm}

\subsubsection{Predictions-as-a-Service (PaaS)}\label{sec:paas}
After the collective model training, the data holders can (i) use the encrypted model for oblivious PaaS \textit{without} decryption, or (ii) if required by the application, they can collectively decrypt the model and reveal it to all the data holders or an external party for semi-oblivious PaaS. In both cases, the querier's evaluation data is encrypted with the data holders' public key and we rely on the collective key-switch functionality of the underlying MHE scheme (Section~\ref{sec:mhe}). In particular, for (i), the querier encrypts $X^q$ with the data holders’ collective key $pk$, the data holders then compute the prediction on the encrypted data using the encrypted model, and switch the encryption key of the result with $\textsf{CKeySwitch}(\mathbf{X}^q_{pk}, qk, [sk_i]$), where $qk$ is the querier's key. Thus, only the querier can decrypt the prediction result. For (ii), the collective decryption is simply a special case of the collective key-switch with the encrypted model as input and no target public key. The rest of the evaluation is same as oblivious PaaS.

\subsection{Challenges Associated with RNN Training}\label{sec:challenges}

We discuss the challenges, as well as several solutions, associated with the training of RNNs in general. We also highlight how these challenges (and their solutions) further increase the difficulty of RNN training under encryption. Section~\ref{sec:crypto} presents the building blocks that \sys employs to alleviate these challenges.

RNNs are inherently challenging to train because of their sequential operations, i.e., the computations performed over the timesteps (the dependency of each network node on the output of its previous one). This phenomenon causes two issues when training RNNs: (i) the training procedure is slow, complex, and hard to parallelize, and (ii) the gradients tend to explode (or vanish) due to long-term dependencies between timesteps in the execution of the back-propagation through time (BBPT) algorithm
~\cite{Pascanu2013,Bengio1994}. 

To mitigate the first challenge, several works suggest accelerating RNN training through data parallelization, i.e., via a distributed approach and/or mini-batch training~\cite{Khomenko,Shi2013Speed,Huang2013}. For example, a master-slave approach where slave machines first compute the gradients and then a master machine aggregates them to perform the global update (Map-Reduce algorithm~\cite{MapReduce_ML}) is commonly proposed. \sys ensures similar efficiency to such distributed training-based approaches, through federated learning that is, by definition, distributed and resembles the master-slave relationship proposed for distributed RNNs. On the other hand, mini-batch training under homomorphic encryption requires a well-suited packing scheme for efficient matrix multiplication and transpose operations. Therefore, in Section~\ref{sec:packing}, we propose a novel multi-dimensional packing scheme that reduces the processing time of a large batch.

The second challenge of exploding or vanishing gradients is addressed by several techniques such as the use of gating mechanisms, e.g., LSTMs~\cite{Hochreiter1997}, along with a possible modification of the gradient propagation~\cite{Kanuparthi}, encoder-decoder approaches~\cite{Cho2014}, gradient clipping~\cite{Pascanu2013}, non-saturating functions~\cite{Chandar2019}, and various recurrent-weight initialization techniques via identity or orthogonal matrices~\cite{Le2015ASW,Henaff}. Employing LSTMs as a solution to the problem of exploding/vanishing gradients brings its own challenges, due to the computational complexity of training LSTMs under homomorphic encryption, and we discuss the costs of such an extension in Section~\ref{sec:complexRNN}. Although various weight-initialization techniques are straightforward for \sys to adopt, we opt for gradient clipping -- a widely used mitigation technique for exploding gradients when training simple RNNs. However, gradient clipping is not easy to compute under encryption as it requires a comparison function that is not homomorphically enabled. Therefore, we present various approximations for the efficient computation of an element-wise clipping function, through different polynomials (Section~\ref{sec:approximations}).


\section{Cryptographic Building Blocks}\label{sec:crypto}
We detail the main cryptographic building blocks that are used in \sys. We describe the packing scheme and the optimized matrix operations including the matrix-multiplication and matrix-transpose in Section~\ref{sec:packing} and the approximated building blocks in Section~\ref{sec:approximations}. We analyze \sys's security in Appendix~\ref{sec:security}.

\begin{figure*}[h]
\centering

    \begin{subfigure}[b]{\textwidth}
    \centering
    \includegraphics[width=0.89\textwidth]{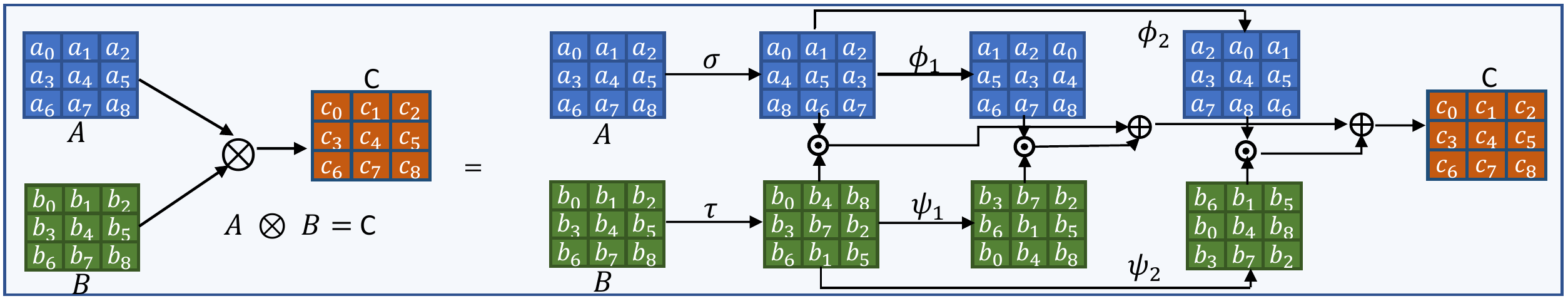} 
    \vspace{-0.3em}
    \captionsetup{width=\textwidth}
    \vspace{-0.3em}
    \caption{Matrix Multiplication Method of
    Jiang et al.~\cite{Jiang2018} for $n=3$.}
    \label{fig:mult}
    \end{subfigure}
    \hfill
    \begin{subfigure}[b]{\textwidth}
        \centering
    \includegraphics[width=0.89\textwidth]{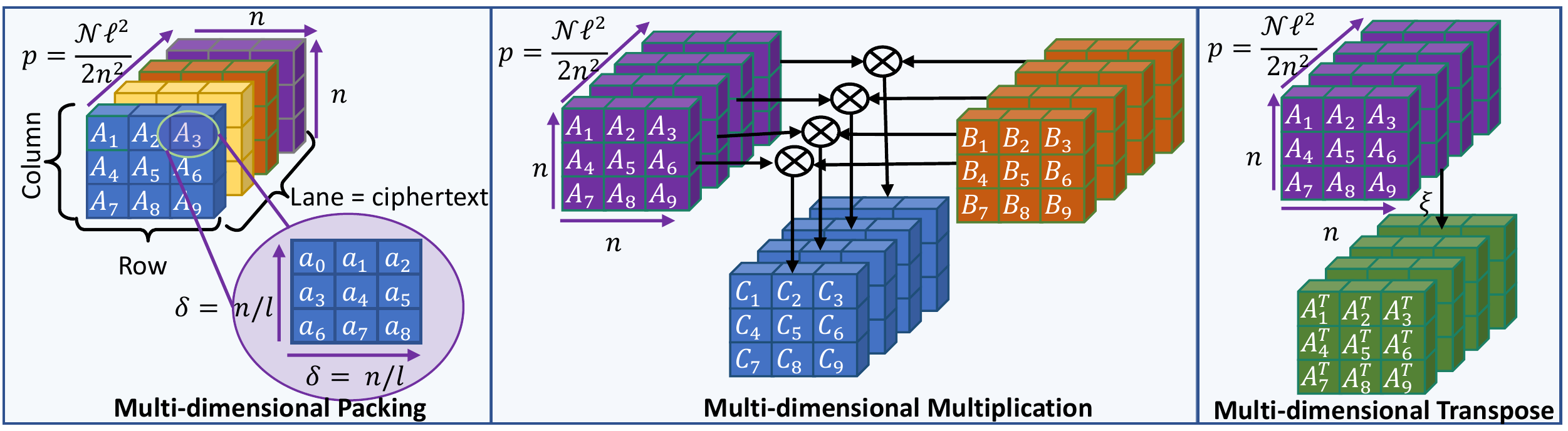}
    \vspace{-0.3em}
    \captionsetup{width=\textwidth}
    \vspace{-0.3em}
    \caption{An overview of our Multi-dimensional Packing scheme, its Multiplication and Transpose operations.}
    \label{fig:multiDim}
    \end{subfigure}
    \vspace{-1em}
    \caption{ Overview of the (a) Matrix Multiplication Protocol by Jiang et al.~\cite{Jiang2018} and (b) our Multi-dimensional Packing Scheme.}
    \vspace{-1em}
    \label{fig:packing}
\end{figure*}

\subsection{Packing and Optimized Matrix Operations}\label{sec:packing}
We present the packing scheme and the optimized matrix operations used for RNN training in \sys. Due to the use of homomorphic encryption (HE), packing has a significant effect on the training performance as through SIMD operations, it enables mini-batch training which alleviates the inherent high training times of RNNs. In addition, matrix operations are the most heavy and frequent operations in RNNs (see Lines 2-5 and 12-17, Algorithm~\ref{algo:rnn}). Thus, optimizations of the packing scheme and matrix operations such as multiplication or transpose, are crucial for \sys's efficient training execution. In this work, we rely on a row-based packing where matrices are decomposed row-wise and packed in one plaintext and we use Jiang et al.'s~\cite{Jiang2018} approach as a baseline for the matrix operations. On top of these matrix operations and row-based packing, we propose and implement a novel multi-dimensional packing scheme that relies on block matrices\footnote{Block matrix is a matrix defined by sub-matrices, called blocks, which enable matrix operations in parallel.} and the matrix operations that are described below, for further optimizations.

\descr{Matrix Multiplication.} The typical methods for matrix multiplication in HE~\cite{Halevi2018,Mishra2021} do not preserve the structure of the matrix in the plaintext slots, i.e., the shape of the computation output is different than that of the inputs. This prevents efficient chaining of matrix operations since after each multiplication an expensive linear transformation is required to convert the output back to an encoding compatible with further operations.

Jiang et al. propose a method for matrix multiplication that preserves the format of row-encoded matrices, enabling the efficient chaining of an arbitrary number of matrix operations~\cite{Jiang2018}. With their approach, a matrix multiplication consumes 3 \textit{levels} per multiplication (2 if one matrix is in plaintext) due to 2 linear transformations and 1 dot product. It enables the packing of multiple matrices in a single ciphertext (assuming the number of matrix-entries is smaller than $\mathcal{N}/2$) and to operate on them in a SIMD manner, which increases efficiency and compensates for the \textit{level} consumption. This feature yields a favorable balance between computation runtime and level consumption for RNN computations. We now describe this format-preserving method for matrix multiplication that is illustrated in Figure~\ref{fig:mult}.
Given $A$ and $B$, matrices of size $n \times n$, the multiplication $C = A \otimes B$ can be evaluated under encryption as the scalar product $\mathbf{C} = \langle\mathbf{\Tilde{A}}, \mathbf{\Tilde{B}}\rangle$, where $\mathbf{\Tilde{A}}$ and $\mathbf{\Tilde{B}}$ are size-$n$ vectors of permutations of $\mathbf{A}$ and $\mathbf{B}$, respectively (or permutations of the original matrices $\mathbf{A}$ and $\mathbf{B}$).
These permutations are represented by the linear transformations $\sigma$ (cyclic rotation of all rows, where the $i$-th row is rotated by $i$ positions), $\tau$ (cyclic rotation of all columns, where the $i$-th column is rotated by $i$ positions), $\phi_i$ (cyclic rotation of all rows by $i$ positions) and $\psi_i$ (cyclic rotation of all columns by $i$ positions).
For example, if $A$ and $B$ are two $4 \times 4$ matrices, then $\mathbf{\Tilde{A}} = \{\phi_0(\sigma(\mathbf{A})), \phi_1(\sigma(\mathbf{A})), \phi_2(\sigma(\mathbf{A})), \phi_3(\sigma(\mathbf{A}))\}$ and $\mathbf{\Tilde{B}} = \{\psi_0(\tau(\mathbf{B})), \psi_1(\tau(\mathbf{B})), \psi_2(\tau(\mathbf{B})), \psi_3(\tau(\mathbf{B}))\}$, and $\mathbf{C} = \langle \mathbf{\Tilde{A}}, \mathbf{\Tilde{B}}\rangle$.
The linear transformations $\sigma$, $\tau$, $\phi_i$ and $\psi_i$ are computationally \textit{cheap} as they can be efficiently parallelized and evaluated with $\mathcal{O}(n)$, $\mathcal{O}(n)$, $\mathcal{O}(1)$, $\mathcal{O}(1)$ homomorphic rotations, respectively. Hence, the matrix multiplication method of Jiang et al. has a complexity of $\mathcal{O}(n)$ rotations for two $n \times n$ matrices. This is smaller than previous approaches, such as the naive approach with $\mathcal{O}(n^{3})$ or the one proposed by Halevi and Shoup~\cite{Halevi2018} with $\mathcal{O}(n^{2})$.

\descr{Matrix Transpose.} Another operation required for RNN training is matrix transpose, as described in Algorithm~\ref{algo:rnn} (Lines 12 and 14). The transpose of a row-packed matrix is straightforward and can be realized with a single linear transform of $2n$ non-zero diagonals~\cite{Jiang2018} ($\xi$), hence, it can be evaluated with $2n$ rotations. This cost can be reduced to $3\sqrt{n}$ by using the baby-step giant-step algorithm proposed in~\cite{Halevi2018} and later improved by Bossuat et al.~\cite{Bossuat2021}. Thus, the cost of a matrix transpose operation is $\mathcal{O}(\sqrt{n})$ for a $n \times n$ matrix.
\begin{algorithm}[t]
\small
\caption{Multi-dimensional Matrix Multiplication}
\label{algo:matrixmultiplicationCipher}
\begin{algorithmic}[1]
\Require $\mathbf{A}$, $\mathbf{B}$ two $\ell\times\ell$ matrices of ciphertexts, each encrypting a sub-matrix of dimension $(n/\ell)\times(n/\ell)$, the linear transformations $\sigma$, $\tau$, $\phi_{i}$ and $\psi_{i}$
\Ensure $\mathbf{C}\leftarrow \mathbf{A}\otimes \mathbf{B}$ 
\For{$i=1 \rightarrow \ell$}
    \For{$j=1 \rightarrow \ell$}
        \State $\mathbf{z} \leftarrow \sigma(\mathbf{B}_{i, j})$ \Comment{Depth 1}
        \State $\mathbf{b}_{i, j} \leftarrow\{\phi_{0}(\mathbf{z}), \dots, \phi_{n/\ell-1}(\mathbf{z})\}$ \Comment{Depth 1}
    \EndFor
\EndFor
\For{$i=1 \rightarrow \ell$}
    \For{$j=1 \rightarrow \ell$} 
        \State $\mathbf{z} \leftarrow \tau(\mathbf{A}_{i, j})$ \Comment{Depth 1}
        \State $\mathbf{a}\leftarrow\{\psi_{0}(\mathbf{z}), \dots, \psi_{n/\ell-1}(\mathbf{z})\}$  \Comment{Depth 1}
        \For{$k=1 \rightarrow \ell$}
            \State $\mathbf{C}_{i, k} \leftarrow \mathbf{C}_{i, k} + \langle \mathbf{a}, \mathbf{b}_{i, j}\rangle$ \Comment{Depth 1}
        \EndFor
    \EndFor
\EndFor
\end{algorithmic}
\end{algorithm}

\descr{Multi-dimensional Packing.} 
We propose a novel packing scheme called multi-dimensional packing to optimize the usage of ciphertext slots, which we illustrate in Figure\ref{fig:multiDim}. Given an $n \times n$ matrix and $\ell$ a divisor of $n$, we split this matrix into $\ell^{2}$ smaller matrices of size $\delta \times \delta$, with $\delta=\tfrac{n}{\ell}$. Each of these 
smaller matrices is stored in a different ciphertext, and we pack in parallel up to $p=\tfrac{\mathcal{N}}{2}\cdot\tfrac{\ell^{2}}{n^{2}}$ block matrices of size $n \times n$ in a set of $\ell^{2}$ ciphertexts. This packing scheme brings several advantages over the plain row-packing approach by Jiang et al: (i) it enables a more efficient packing of non-square matrices with a worst case of extra space used reduced from $n(n-1)$ to $\tfrac{n}{\ell}(\tfrac{n}{\ell} - 1)$,
(ii) it provides a better amortized multiplication complexity as it packs up to $\tfrac{\mathcal{N}}{2}\cdot\tfrac{\ell^{2}}{n^{2}}$ matrices (instead of $\tfrac{\mathcal{N}}{2n^{2}}$) in parallel with $\mathcal{O}(n\ell)$ complexity for the multiplication, but amortizes to $\mathcal{O}(n/\ell)$ per matrix (instead of $\mathcal{O}(n)$), (iii) it reduces the complexity of the transpose operation from $\mathcal{O}(\sqrt{n})$ to $\mathcal{O}(\sqrt{n/\ell})$ (only the $\xi$ permutation needs to be evaluated on the sub-matrices and permutations between ciphertexts are free), and (iv) it enables more efficient matrix slot manipulations as rows of sub-matrices can be individually moved and/or rotated.
We further minimize the number of redundant homomorphic operations by leveraging the techniques used in the \textit{double-hoisting} baby-step giant-step algorithm~\cite{Bossuat2021} to evaluate linear transforms and delay ciphertext relinearization. We describe the protocol for multi-dimensional matrix multiplication in Algorithm~\ref{algo:matrixmultiplicationCipher}. We also present guidelines for configuring $\delta$ and other cryptographic parameters in Appendix~\ref{sec:paramSelection}.

\sys employs the multi-dimensional packing for efficient mini-batch training under encryption. In particular, we distribute the RNN input batch along the ``third" dimension (see Figure~\ref{fig:multiDim}), by expressing the batch as $p$ parallel matrices (as if computing an ensemble of $p$ parallel RNNs). We can increase the batch size (up to $p*n/\ell$) at no additional cost as it fits into the same ciphertext. To enable further operations, i.e., to make the input batch compatible with the weight matrices, we replicate the latter across the third dimension, so that the weight matrices are aligned with their parallel slices from the input batch. The computation of the FP and BP for all the timesteps results in a parallel slicing of the gradients across the third dimension. These gradients are then aggregated (with $\log(p)$ rotations to perform an inner sum operation), to ensure consistency among the parallel instances. Note that for batches that fit into one ciphertext, our multi-dimensional packing is equivalent to an optimized version (via block-matrices) of Jiang et al.'s~\cite{Jiang2018} approach. However, for bigger dimensions, where the matrix is split over more than one ciphertext, multi-dimensional packing yields better amortized matrix multiplication cost (see Section~\ref{sec:microbenchmarks}).

\subsection{Approximated RNN Building Blocks}\label{sec:approximations}
To compute any non-linear activation or clipping function under CKKS encryption, \sys relies on polynomial approximations employing the least-squares or Minimax methods. The least-squares method finds the optimal polynomial that minimizes the squared error between the real function and its approximation over an interval, whereas Minimax minimizes the corresponding maximum error. Admittedly, the maximum error with Minimax decreases as the approximation degree ($\mathfrak{p}$) increases or as the range of the interval shrinks. However, note that with larger $\mathfrak{p}$, the polynomial evaluation becomes more expensive (with a scale of $O(\log({\mathfrak{p}}))$) and smaller interval ranges are not always possible due to the need of accommodating the range of the recurrent neural network outputs. We remark that choosing $\mathfrak{p}$ and the interval of approximations is a non-trivial task under privacy constraints. Yet, these can be determined by synthetic datasets or based on data distribution-based heuristics such as computing the minimum, the maximum, and the mean of feature vector means per dataset~\cite{cryptoDL}. 

\descr{Activation Functions.} A neural network is a pipeline of layers composed of neurons on which linear and non-linear transformations (activations) are applied. Typical activation functions include $\textsf{Sigmoid}$ ($\varphi(x)=$ \sigmoid), $\textsf{ReLU}$ ($\varphi(x)=$\relu), $\textsf{Tanh}$ ($\varphi(x)=$ \tanhh), etc., that are not computable under encryption as they comprise non-linear operations (e.g., comparison and exponentiation). To overcome this issue, previous works either change the function to a linear or square one~\cite{Maya2020,CryptoNets}, or rely on various approximation techniques employing the least-squares method~\cite{spindle,poseidon}, polynomial splines~\cite{MiniONN}, Legendre~\cite{cryptoDL} or Chebyshev polynomials~\cite{lattigo,cryptoDL}. For RNNs, the choice of activation function is critical due to exploding gradients problem (i.e., the function should be bounded) and commonly used functions are $\textsf{Tanh}$ and $\textsf{ReLU}$ (that can approximated as $\textsf{SoftPlus}$, i.e., $\varphi(x)=$\softplus~\cite{poseidon}). Thus, similar to prior work, we rely on polynomial approximations 
to enable the execution of these activation functions.

\descr{Gradient Clipping.} As described in Section~\ref{sec:challenges}, when the weights have a large norm, the gradients accumulated over multiple steps of the BBPT algorithm can grow exponentially; gradient clipping is one of the well-established techniques to mitigate this issue. Thus, in \sys, the \textit{aggregation server} applies clipping during the Model Update Phase (see Section~\ref{sec:solutionOverview}). Pascanu et al.~\cite{Pascanu2013} propose clipping the gradients based on their infinity-norm,
yet applying directly their function under encryption is challenging due to the norm calculation that comprises comparison operations. Thus, inspired from their norm-clipping, we propose a gradient clipping strategy that restricts the infinity-norm of the gradients to a closed interval whose limits are defined by a threshold $|m|$. We apply the following function element-wise to the accumulated gradient vector $x$:
\vspace{-0.3em}
\[\textsf{Clip}(x, m) = \begin{cases} 
  -m & x\leq -m \\
  x & -m < x <m \\
  m  & x\geq m
\end{cases}
\]

\noindent $\textsf{Clip}(x, m)$ is a baseline as this function is also difficult to practically approximate due to its non-smoothness near the clipping interval limits $[-m, m]$. Therefore, we introduce two \textit{softer} approximations for clipping gradients. The first one is based on $\textsf{Tanh}$, which is very similar to $\textsf{Clip}$ except for the clipping interval limits $-m$ and $m$:

\vspace{-0.3em}
\[\textsf{TanhClip}(x,m) = m*tanh\left(\frac{x}{m}\right)\]


\noindent The second one is based on $\textsf{ReLU}$, i.e., $\textsf{Clip}(x,m) = x + \textsf{ReLU}(-(x+m)) - \textsf{ReLU}(x-m)$.
Since \sys softly approximates the $\textsf{ReLU}$ function with \textsf{SoftPlus}, $\textsf{SoftClip}$ is defined accordingly as $\textsf{SoftClip}(x, m) = x+\textsf{SoftPlus}(-(x+m)) - \textsf{SoftPlus}(x-m)$, or:
\vspace{-0.3em}
    \[\textsf{SoftClip}(x, m)=x+\ln{\frac{1+e^{-(x+m)}}{1+e^{(x-m)}}}\]

\noindent Overall, \sys enables the approximated execution of both $\textsf{TanhClip}$ and $\textsf{SoftClip}$. We evaluate the performance of both clipping approximations and provide the plots of the approximated curves (by Minimax) and their errors for polynomial degrees of $\mathfrak{p}=5$ and $\mathfrak{p}=15$ in Appendix~\ref{sec:approxCurve}. We observe that both functions achieve a similar average error over the approximation interval, yet $\textsf{SoftClip}$ is slightly better than $\textsf{TanhClip}$ for higher degrees in terms of error and behavior around the limits of the approximation interval.

\section{Extension to More Complex RNNs}\label{sec:complexRNN}

So far, we have described \sys's workflow with a simple RNN architecture, yet, \sys can be extended to more complex ones such as GRUs and LSTMs. \sys's main building blocks, i.e., the approximated activation/clipping functions, the multi-dimensional packing, the matrix multiplication and transpose operations are sufficient to implement such complex RNNs. As the multi-dimensional packing preserves the structure of the matrices in the plaintext slots after multiplication and transpose operations, any pipeline can be built from our blocks. RNNs such as GRUs or LSTMs, require changes only on the Local Computation Phase of \sys. We now briefly explain how to extend \sys to GRUs and LSTMs.

\descr{GRU.} We summarize the protocol for the GRU architecture in Algorithm~\ref{algorithm:gru}, Appendix~\ref{sec:gru_algo}. Such an RNN has the same workflow as a simple RNN but with different operations inside the GRU unit. A typical GRU comprises an Update Gate (R-Gate) that decides how much information from previous timesteps is needed for future ones, a Reset Gate (Z-Gate) that decides how much information to forget, and a Candidate activation (N-Gate) similar to the hidden state of an RNN unit. Other GRU variants modify these gates by computing only (or excluding) the biases~\cite{Heck2017,Rahul2017}. Both Z- and R-Gates include an activation function that is typically a $\textsf{Sigmoid}$, $\varrho=\sigmoid$, and the hidden state comprises the $\textsf{Tanh}$ activation function, $\varphi=\tanhh$. Then, the pipeline for the forward pass (FP) of a GRU is similar to a simple RNN and the backpropagation (BP) follows the BBPT with the chain rule for the calculation of the derivatives and the loss (see Algorithm~\ref{algorithm:gru}). In more detail, the FP (Lines 2-10) comprises encrypted matrix multiplications, additions, and activation functions, thus, the packing scheme of \sys, the matrix multiplication and the approximated building blocks can be directly applied. Similarly, the BP comprises the same operations as the FP but additionally requires multiple executions of the transpose, which is also supported by \sys (Section~\ref{sec:packing}). 

\descr{LSTM.} LSTM includes a Forget Gate (f), an Input Gate (i), a Cell State (c), and an Output Gate (o). The hidden state is calculated as:
\begin{equation*}
\begin{split}
f_t &= \varrho( W_f\times x_t + U_f \times h_{t-1} +b_f) \quad \quad\, \vartriangleright \text{Forget Gate}\\
i_t &= \varrho( W_i\times x_t + U_i \times h_{t-1} +b_i)\quad \quad \;\;\: \vartriangleright \text{Input (Update) Gate}\\
o_t &= \varrho( W_o\times x_t + U_o \times h_{t-1} +b_o)\quad \quad \;\vartriangleright \text{Output Gate}\\
c_a &= \varphi( W_c\times x_t + U_c \times h_{t-1} +b_c)\quad \quad \;\:\vartriangleright \text{Cell Input Activation}\\
c_t &= f_t \odot c_{t-1} + i_t \odot c_a \qquad \qquad \qquad\;\:\vartriangleright \text{Cell State}\\
h_t &= o_t \odot c_{t-1} + \varphi(c_t)  \qquad \qquad \qquad\quad\vartriangleright \text{Hidden State}\\
\end{split}
\label{eq:lstm}
\end{equation*}
where $U_\cdot$ and $W_\cdot$ denote the weight matrices, $b_\cdot$ the bias vector, with the respective gate names as sub-indices ($\cdot$). As in GRUs, the MHE execution of LSTM requires the matrix multiplication, element-wise multiplication, addition, and activation functions (approximated via our building blocks). We note that the element-wise multiplication with multi-dimensional packing is straight-forward and requires $\ell^2$ ciphertext multiplications. Due to the nature of the operations, the BP for LSTM training is similar to GRU-BP and requires the exact same operations of the forward pass with the addition of the transpose operation which is supported by \sys.

Overall, different RNN architectures composed of the aforementioned operations can be enabled using \sys's building blocks. Yet, it is worth mentioning that implementing these models under MHE is non-trivial and will increase \sys's computation and communication overhead; its experimental evaluation on such complex RNN architectures is a direction for future work. Here, we roughly estimate the overhead incurred by GRUs and LSTMs. For the forward pass, a GRU requires $3\times$ the operations (i.e., 2 multiplication and additions followed by an activation) of Lines 3-4, Algorithm~\ref{algo:rnn} and element-wise multiplications which are negligible compared to matrix multiplications. Then, the prediction (Line 5, Algorithm~\ref{algo:rnn}) remains the same. Due to the chain rule, the same rationale holds for the BP. Similarly, LSTMs require $4\times$ the operations of Lines 3-4 of Algorithm~\ref{algo:rnn} followed by $3\times$ more element-wise multiplications, but with one less matrix multiplication for the prediction step. Thus, we estimate that implementing GRUs and LSTMs \textit{without parallelization} will result in a $3\times$ -- $4\times$ slower local computation phase than simple RNNs. Finally, as the number of GRU and LSTM weight matrices are $3\times$ and $4\times$ bigger than simple RNNs, estimates apply for the communication overhead incurred by collective bootstrapping and the FL workflow.
\section{System Evaluation}\label{sec:evaluation}

We analyze \sys's theoretical complexity in Section~\ref{sec:complexity}. Then, we present our experimental setup in Section~\ref{sec:experimentSetup}. We evaluate \sys's model performance, scalability with different parameters, and runtime in Sections~\ref{sec:modelPerformance},~\ref{sec:scalability}, and~\ref{sec:runtimePerformance}, respectively. Finally, we 
compare \sys with prior work in Section~\ref{sec:microbenchmarks}.

\subsection{Complexity Analysis}\label{sec:complexity}

We summarize here \sys's complexity by taking into account the memory usage per data holder, as well as its communication and computational costs for an Elman network (Algorithm~\ref{algo:rnn}). Here, we discuss the dominating terms that affect the memory, communication, and computation costs and we provide a more detailed analysis with an example network configuration in Appendix~\ref{sec:complexityDetailed}. 
The memory usage per data holder depends on the number of input plaintexts, activation ciphertexts (e.g., $\mathbf{h}_t$ in Algorithm~\ref{algo:rnn}), weight ciphertexts (e.g., $\mathbf{U},\mathbf{V},\mathbf{W}$), and gradient ciphertexts (e.g., $\mathbf{\nabla U},\mathbf{\nabla W},\mathbf{\nabla V}$).
Subsequently, the number of ciphertexts storing the matrix and vectors required for \sys's execution has a direct impact on the communication cost. Recall that during RNN training, communication among the data holders is triggered by: (a) the collective bootstrapping ($\textsf{CBootstrap}(\cdot)$) operation to refresh the ciphertexts, and (b) the federated learning (FL) workflow, where data holders communicate their local gradients (Aggregate Phase). Both (a) and (b) incur costs that depend on the number and size of the weight/gradient ciphertexts (see Appendix~\ref{sec:complexityDetailed} for details).

\sys's computational cost per data holder and per local iteration depends on the number of $\textsf{CBootstrap}(\cdot)$ operations (which are $\sim$2 orders of magnitude slower than a homomorphic addition/multiplication), the degree $\mathfrak{p}$ of the polynomial approximations, and the cost of the linear transformations (i.e., $\sigma$, $\tau$, $\psi_i$ and $\phi_i$) required for the matrix multiplication.
Indeed, these linear transformations dominate the complexity of the matrix multiplication (see Algorithm~\ref{algo:matrixmultiplicationCipher}); the rest of the algorithm comprises only multiplications and additions, which are negligible in comparison. Overall, \sys's RNN training protocol requires several multiplications with the weight matrices ($\mathbf{W}, \mathbf{U},\mathbf{V}$) and their transposes ($\mathbf{V}^T, \mathbf{W}^T$). As these matrices are not updated through the local iteration \textit{per} timestep, we can pre-compute their linear transformations at the beginning of each iteration and re-use them for all timesteps. For the network structure in Algorithm~\ref{algo:rnn}, this reduces the cost to $\mathcal{O}(6\delta)$ from $\mathcal{O}(6T\delta)$ per local iteration, at a memory cost of $\times\mathbb{d}$ per matrix.

\subsection{Experimental Setup}\label{sec:experimentSetup}
We present the datasets used for model performance evaluation. We describe the data distribution, RNN configuration, and security parameters. We provide the implementation details in Appendix~\ref{sec:Implementation}.



\descr{Datasets and Tasks.} We employ the Hourly Energy Consumption (HEC)~\cite{hecKaggle}, Stock Prices (Stock)~\cite{stocks}, and Inflation datasets~\cite{inflation}, for time-series forecasting tasks. HEC contains the hourly energy consumption of several electricity distribution companies during a specific time period~\cite{hecKaggle}. The forecasting task is to use early sequences of energy consumption and predict the future value for sequences of length $T$. The Stock dataset contains historical daily stock statistics for 10 companies~\cite{stocks}. The task on this dataset is to predict the average open-high-low-closing (OHLC) price of the next day, given the OHLC of the previous $T$ days. The Inflation dataset contains quarterly inflation rates from 40 countries and the task is to predict the inflation rate of the next quarter given the previous $T$ quarters. We also employ the Breast Cancer Wisconsin dataset (BCW)~\cite{bcw} that contains benign and malignant breast cancer samples for classification. More details about datasets can be found in Appendix~\ref{sec:appendixDataset}. For each experimental setting, we split the dataset into training ($80\%$) and test ($20\%$) sets. For the \sys's scalability and microbenchmark experiments, we use synthetic datasets.

\descr{Dataset Distribution.} For the model performance experiments (Section~\ref{sec:modelPerformance}), we split the data among $N{=}10$ data holders and evaluate two different settings: (i) \textit{Even (E)} distribution, where we uniformly distribute the dataset to $10$ data holders, and (ii) \textit{Imbalanced (I)} distribution where we simulate a heterogeneous setting (see Appendix~\ref{sec:appendixDataset}). We annotate the results per setting as "DatasetName-\textit{Type}T" where \textit{Type}T denotes the type of the data distribution with the sequence length $T$, e.g., HEC-\textit{I}10, denotes imbalanced data distribution and $T{=}10$ on the HEC dataset.

\descr{RNN Configuration.} For the HEC and Stocks model performance experiments, we use an Elman network with a local batch size of $b{=}256$, a hidden dimension of $h{=}32$, a learning rate of $\eta{=}0.1$, and varying timesteps $T{=}[5-20]$, and we train the RNN for $e{=}1,000$ and $e{=}300$ global iterations, resp. For the Inflation dataset, we employ a Jordan network that is widely used for financial forecasting tasks~\cite{Tea2021,Moshiri1999,Tea2019,Hardinata2018BankruptcyPB,Gabriel2012,DAMATO2022127158} with parameters $b{=}128$, $h{=}16$, $\eta{=}0.1$, $T{=}8$, and $e{=}500$. For BCW, we evaluate an Elman network used for breast cancer classification~\cite{Lili2020,Chunekar2009,Zakia2021,Nagalakshmi2022} with $b{=}32$, $h{=}64$, $\eta{=}0.2$, $T{=}9$, and $e{=}400$. For all experiments, the data holders perform one local iteration and we use the same approximation parameters: $\textsf{SoftClip}(x, m)$ with a clipping threshold of $|m|{=}5$, approximated with a $\mathfrak{p}{=}7$ polynomial over the interval $[-60, 60]$ as a result of a preliminary evaluations on the datasets. To expedite the model performance experiments (Section~\ref{sec:modelPerformance}), we simulate \sys's fully-encrypted training pipeline in plaintext by using the approximated activation and clipping functions, and a fixed-precision.

\descr{Security Parameters.} Unless otherwise stated, we set the degree of the cyclotomic polynomial to $\mathcal{N}{=}2^{14}$ and the modulus of the keys to $Q{\approx}2^{438}$. These parameters yield 128-bit security according to the homomorphic encryption standard~\cite{HEStandardPaper}. We set the plaintext scale to $2^{31}$, and the initial ciphertext level is $\mathcal{L}{=}9$.

\begin{table}[t]
\centering
\small
\setlength{\tabcolsep}{2.6pt}
\begin{tabular}{l cccc}
\toprule
 \multirow{2}{*}{Dataset}  & \multicolumn{4}{c}{Model Performance (MAE | $R^2$ or Acc ) }  \\
\cmidrule(lr){2-5}
& \multicolumn{1}{c}{\textbf{L}} & \multicolumn{1}{c}{\textbf{C}} & \multicolumn{1}{c}{\textbf{FL}} & \multicolumn{1}{c}{\textbf{\sys}} \\
\midrule
\textbf{HEC-\textit{E}10} & 0.066 | 0.749 & 0.014 | 0.987 & 0.024 | 0.965 &  0.026 | 0.957  \\

\textbf{HEC-\textit{I}10} & 0.081 | 0.625 & 0.014 | 0.987 & 0.025 | 0.962 & 0.027 | 0.955 \\
\midrule
\textbf{HEC-\textit{E}20} & 0.066 | 0.749 & 0.014 | 0.986 & 0.024 | 0.964 & 0.026 | 0.957\\

\textbf{HEC-\textit{I}20} & 0.081 | 0.626 & 0.014 | 0.986 & 0.025 | 0.962 & 0.027 | 0.954 \\
\midrule

\textbf{Stock-\textit{E}5} & 0.023 | 0.966 & 0.008 | 0.996 & 0.020 | 0.982 & 0.017 | 0.987  \\

\textbf{Stock-\textit{I}5} &  0.024 | 0.976 & 0.008 | 0.996 &0.012 | 0.992  & 0.012 | 0.992 \\
\midrule
\textbf{Inflation-\textit{E}8} & 0.067 | 0.778& 0.067 | 0.775 & 0.067 | 0.774  &  0.067 | 0.771  \\
\textbf{Inflation-\textit{I}8} & 0.103 | 0.579 &  0.066 | 0.777  & 0.066 | 0.777  &  0.067 | 0.770 \\
\midrule
\textbf{BCW-\textit{E}9}&  0.888 &  0.936  & 0.936  &  0.929  \\
\textbf{BCW-\textit{I}9}& 0.880 &  0.936  & 0.900  &  0.914  \\
\bottomrule
\end{tabular}
\captionsetup{width=\linewidth}
\caption{\sys's model performance in various settings ($N{=}10$) compared to three baselines: local training (\textbf{L}), centralized (\textbf{C}), and federated learning (\textbf{FL}).}
\label{table:performance}
\end{table}

\begin{figure*}[t]
	\centering
	\begin{subfigure}[t]{0.3\textwidth}
		\centering
		\includegraphics[width=0.8\columnwidth]
		{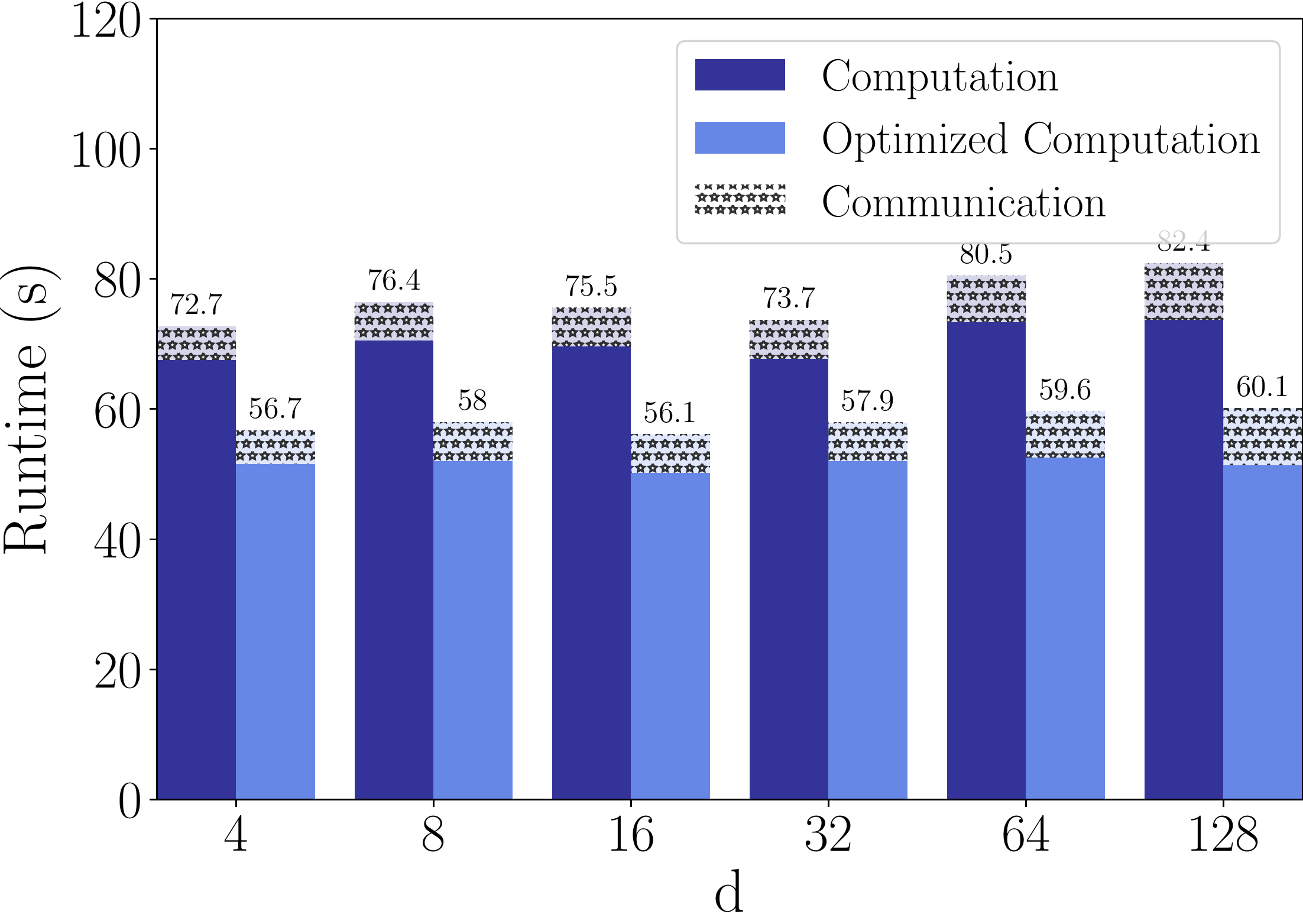}
				\vspace{-0.3em}

		\captionsetup{width=0.975\linewidth}

		\caption{Increasing number of features ($d$).}
		\label{fig:scalingFeature}
	\end{subfigure}
		\begin{subfigure}[t]{0.3\textwidth}
		\centering
		\includegraphics[width=0.8\columnwidth]
		{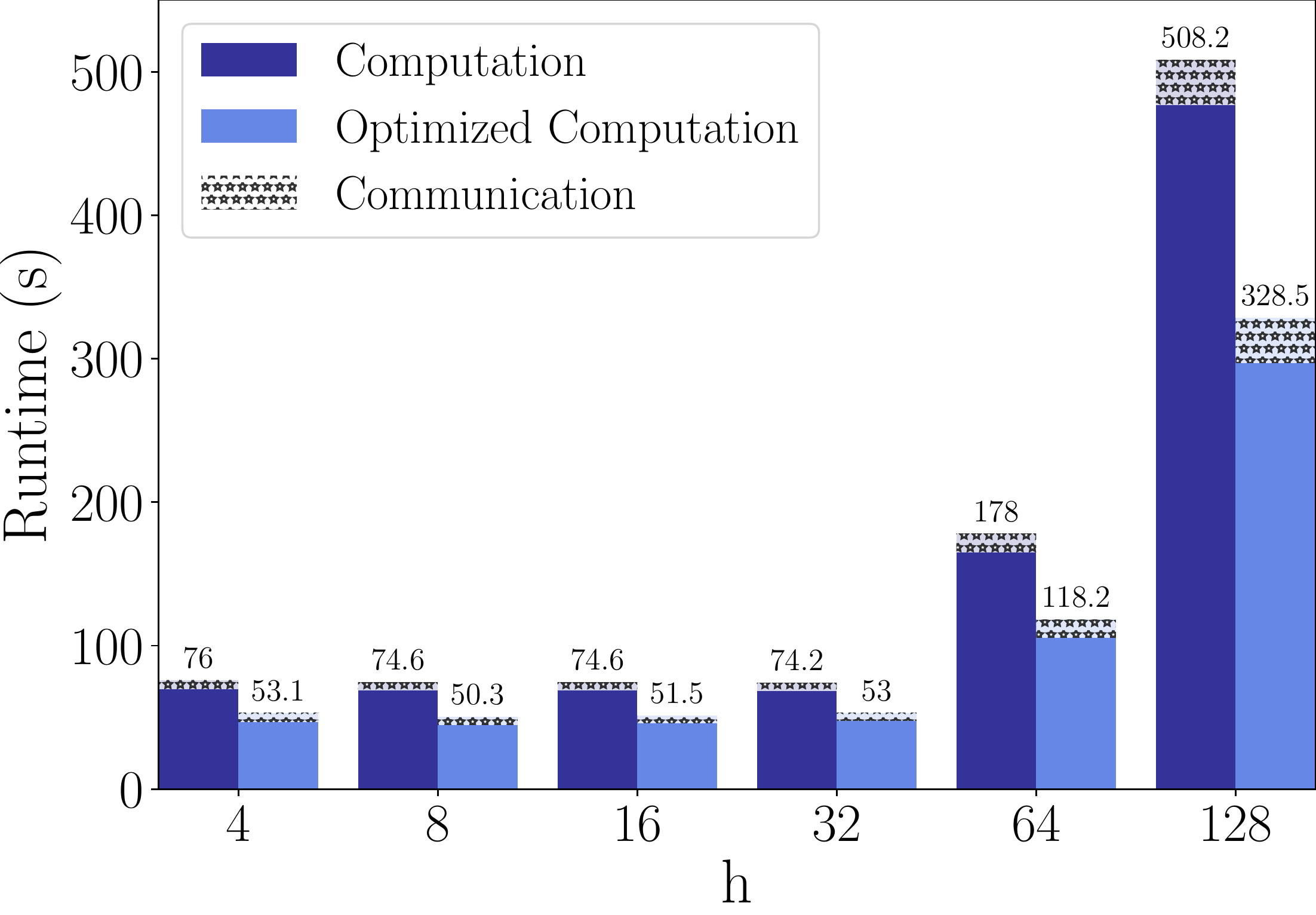}
				\vspace{-0.3em}

		\captionsetup{width=0.975\linewidth}

		\caption{Increasing hidden dimension ($h$).}
		\label{fig:scalingHidden}
	\end{subfigure}
	 \vskip\baselineskip
	 		\vspace{-1em}

	\begin{subfigure}[t]{0.3\textwidth}
		\centering
		\includegraphics[width=0.8\columnwidth]
		{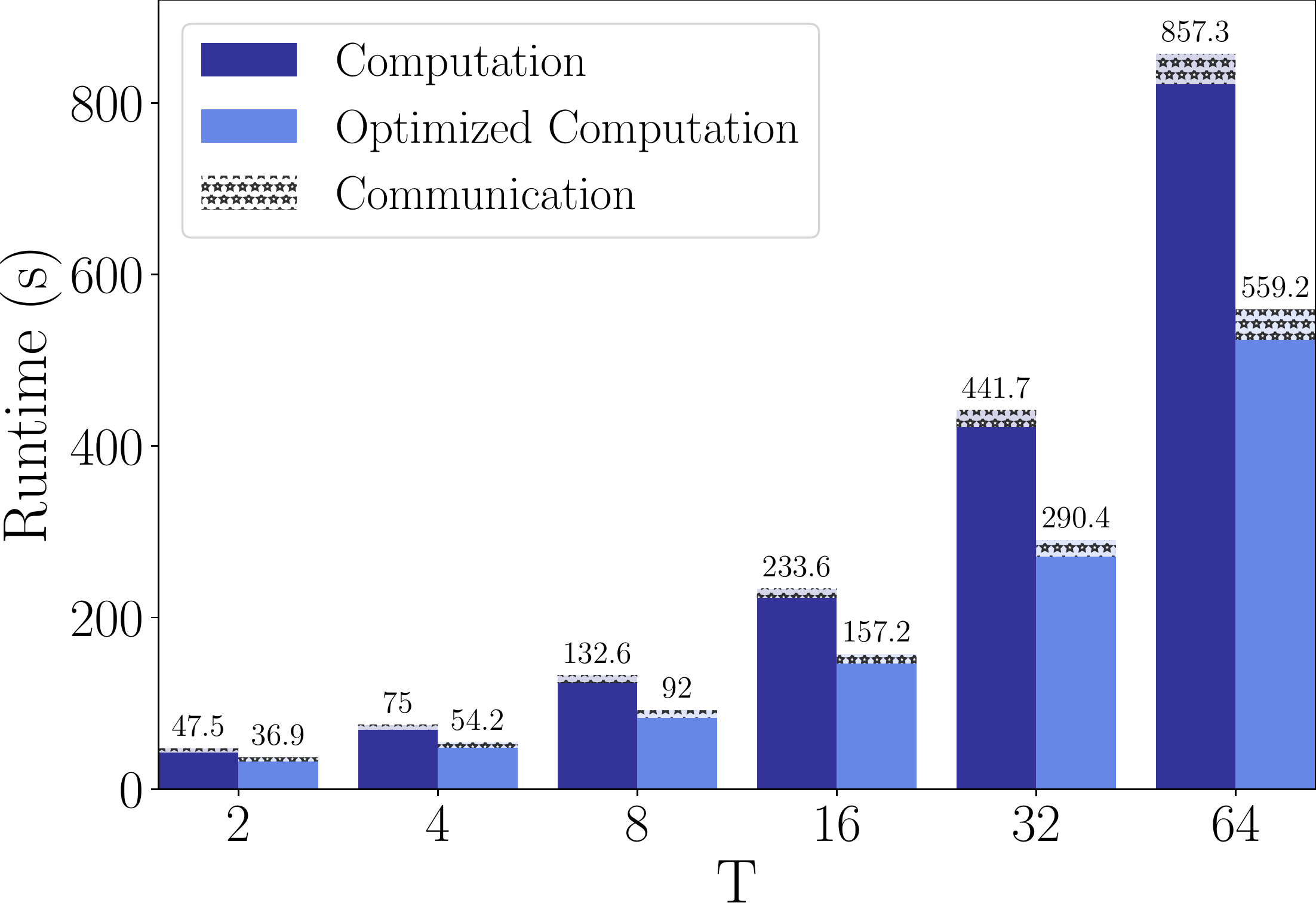}
		\vspace{-0.3em}
		\captionsetup{width=0.975\linewidth}

		\caption{Increasing number of timesteps ($T$).}
		\label{fig:scalingTimesteps}
	\end{subfigure}
	\begin{subfigure}[t]{0.3\textwidth}
		\centering
		\includegraphics[width=0.8\columnwidth]
		{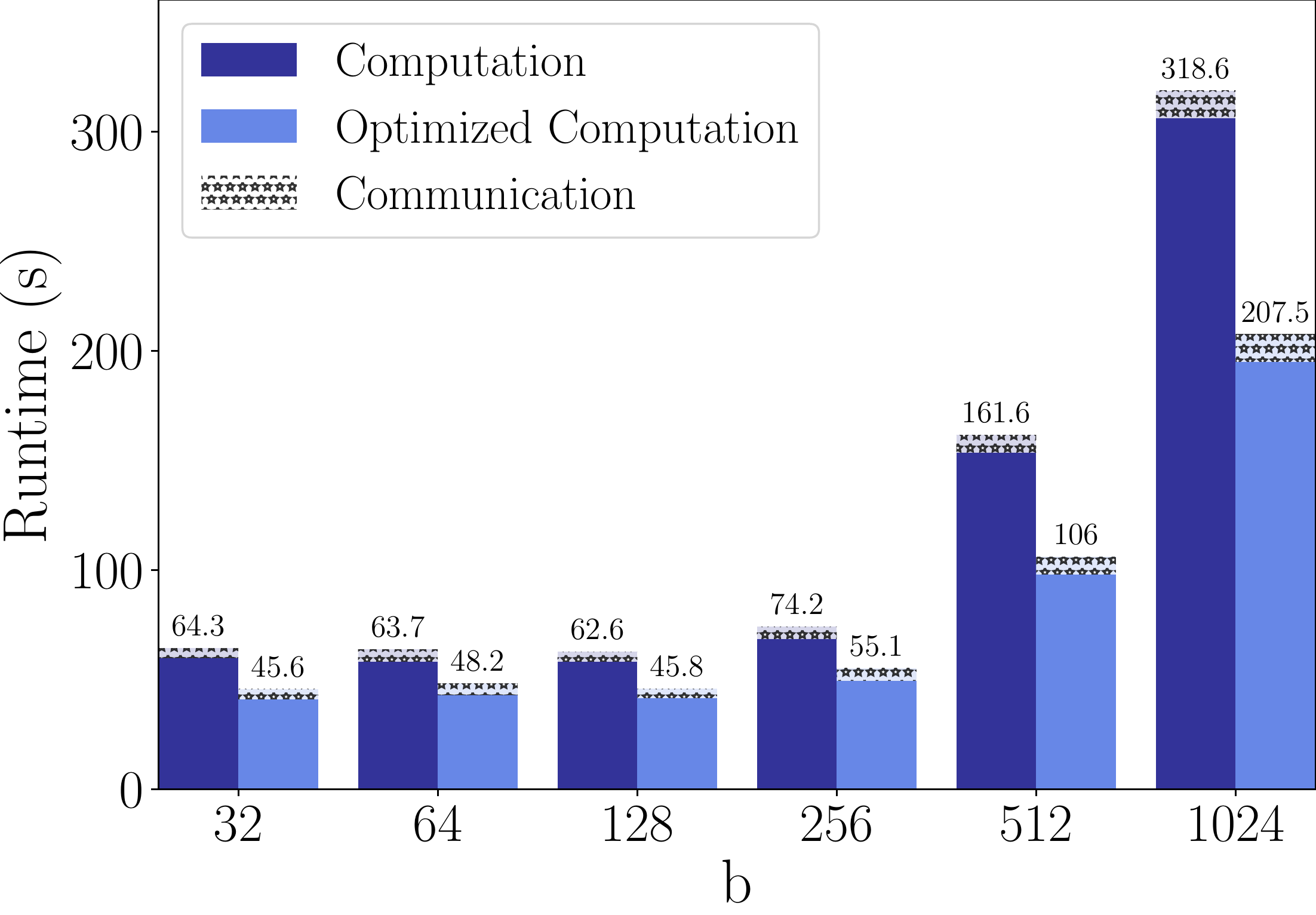}
		\vspace{-0.3em}
		\captionsetup{width=0.975\linewidth}

		\caption{Increasing batch size ($b$).}
		\label{fig:scalingBatch}
	\end{subfigure}
		\begin{subfigure}[t]{0.3\textwidth}
		\centering
		\includegraphics[width=0.8\columnwidth]
		{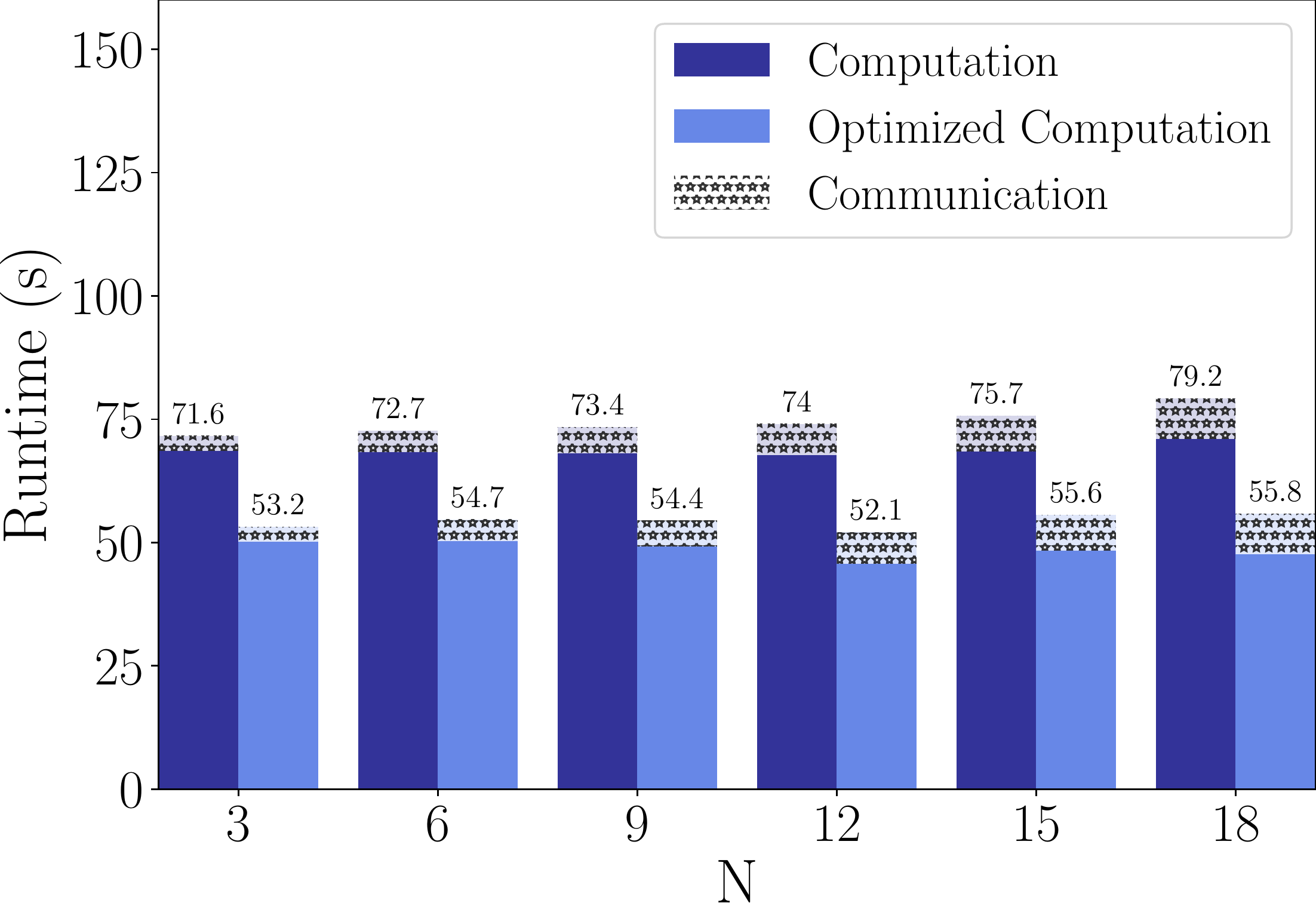}
		\vspace{-0.5em}
					\captionsetup{width=0.975\linewidth}
		\caption{Increasing number of data holders ($N$).}
		\label{fig:scalingN}
\end{subfigure}
	\captionsetup{width=\linewidth}
	\vspace{-0.2em}
	\caption{\sys's training execution times for computation and communication of one global iteration with increasing \textbf{(a)} number of features, \textbf{(b)} hidden dimension, \textbf{(c)} timesteps, \textbf{(d)} batch size, and \textbf{(e)} number of data holders. The default parameters are $h{=}d{=}T{=}4,b{=}256,\delta{=}32$, $N{=}10$ (we vary one of them in the corresponding experiment while keeping the others fixed).}
	\label{fig:scaling}
\end{figure*}
\subsection{Model Performance}\label{sec:modelPerformance}
Table~\ref{table:performance} shows \sys's model performance results on the various datasets and settings. For each setting, we report the mean absolute error (MAE) and $R^2$-scores for forecasting tasks and accuracy (Acc) for the classification one. For comparison, the table displays the performance of three baselines: (a) \textbf{L} stands for local training, where each data holder trains its own local model with the original network (original activation functions and clipping), i.e., without taking advantage of the other parties' data, (b) \textbf{C} stands for centralized training, where all the data is collected to a central server and trained with the original network, and (c) \textbf{FL} stands for original FL, where the data is distributed among 10 data holders and the learning is performed without any privacy mechanism. Lastly, \sys for privacy-preserving FL. The baseline \textbf{L} shows the performance gain of collaborative training (note that we report the average MAE/$R^2$/Acc across the data holders). Whereas, the baselines \textbf{C} and \textbf{FL} enable us to evaluate \sys's performance loss due to collective learning, the approximation of the activation/clipping functions, and the encryption.
In all settings, we observe that \sys achieves a model performance comparable to non-private baselines with at most $0.03$ ($0.01$) difference in $R^2$-score compared to a centralized (federated learning) solution for the three forecasting tasks and at most $1\%$ accuracy difference for the classification task. We also observe that there is always a performance gain, compared to \textbf{L} (except for the even distribution of the Inflation dataset where the local data already captures the global insights). The gain in terms of $R^2$-score for the Stock dataset is between $0.01-0.02$, whereas it is between $0.21-0.33$ for the more complex HEC dataset. More importantly, for an imbalanced distribution, i.e., HEC-\textit{I}10 or HEC-\textit{I}20, the gain from collaborative learning is more substantial ($\sim$0.33 increase in $R^2$-score). We observe a similar trend on the MAE, justifying the use of privacy-preserving collaborative learning. Similarly, \sys yields a $\sim$3$\%$ accuracy gain on BCW compared to \textbf{L}.

\subsection{Scalability Analysis}\label{sec:scalability}
We evaluate \sys's scalability by varying the number of features ($d$), the hidden dimension ($h$), the timesteps ($T$), and the batch size ($b$), and by experimenting with the number of data holders ($N$). Our default parameters are $h{=}d{=}T{=}4$, $b{=}256$, $\delta{=}32$, $N{=}10$, and we vary the parameter under analysis. The calculated runtime includes average per-data holder Local Computation, as well as the Aggregate, and Model-Update Phases. Figure~\ref{fig:scaling} shows the time spent by \sys (and its optimized version with pre-computed linear transforms, described in Section~\ref{sec:complexity}) on computation and on communication during a global iteration. The communication indicates the total time spent for the collective bootstrapping operations ($\textsf{CBootstrap}(\cdot)$) and the collective aggregation.
\begin{table}[h]
\centering
\small
\setlength{\tabcolsep}{2.7pt}
\renewcommand{\arraystretch}{0.8}
\begin{tabular}{ccccc}
\toprule
 & \multicolumn{4}{c}{Runtime (s)} \\ \cmidrule{2-5}
\begin{tabular}[c]{@{}c@{}}Structure\\ ($b$,$d$,$h$,$T$)\end{tabular} & Training & Training (Opt) & Prediction & Prediction (Opt) \\
\midrule
(256, 4, 32, 4) & 72.65 & 56.68  & 27.61 & 18.10\\
(256, 32, 32, 4) & 73.66 &  57.94  & 28.06 & 18.55\\
(256, 64, 32, 4) & 80.48 & 59.63 & 31.04 & 20.53\\
(256, 4, 16, 4) & 74.56 & 51.45 & 29.28 & 15.50  \\
(256, 4, 32, 4) & 74.19 & 53.04 & 71.92 & 16.31 \\
(256, 4, 64, 4) & 178.01 & 118.17 & 225.34 &  56.41\\
\bottomrule
\end{tabular}
\captionsetup{width=\linewidth}
\caption{\sys's training and prediction runtime for several RNNs with $N{=}10$. Training runtimes are for one global iteration and for oblivious predictions on $256$ samples, resp. Opt stands for optimized execution.}
\label{table:runtime}
\end{table}
Overall, we observe that the optimized computation is always more efficient in terms of runtime at a memory cost of $32\times$ more ciphertexts per weight matrix (see Section~\ref{sec:complexity}). Moreover, Figures~\ref{fig:scalingFeature},~\ref{fig:scalingHidden} and~\ref{fig:scalingBatch} demonstrate that \sys's computation time remains constant, with increasing the respective parameters $\{d,h,b\}$ when ${d,h,b}<\delta$. This is due to the SIMD operations;
as for these settings the matrix entries fit into one ciphertext, \sys enables processing larger batches or larger networks at no additional cost. Figure~\ref{fig:scalingFeature} shows that \sys scales sub-linearly with $d$ and that the effect of $d$ is almost negligible in runtime, because each data holder's data remains in clear through the training and the plaintext operations are negligible, compared to ciphertext ones. Increasing $d$ has a direct effect on the matrix multiplication ($x_t \times \mathbf{U}$) of the first layer, as $x_t$ is a plaintext whose size depends on $d$. As a result, the effect of increasing $d$ is not observable; the ciphertext multiplications ($ \mathbf{h}_{t-1} \times \mathbf{W} $) dominate the runtime. The communication overhead also scales sub-linearly; as increasing $d$ increases the number of ciphertexts to be bootstrapped.
Figure~\ref{fig:scalingHidden} shows that \sys scales sub-quadratically, with increasing $h$ when $\delta{=}32$; while increasing $h$ quadratically increases the number of ciphertexts for the weight matrix $\mathbf{W}$, this affects the runtime of a few matrix multiplications only (see Section~\ref{sec:complexity}). This trend holds also for the communication due to an increased number of ciphertexts that are bootstrapped by the aggregation server. Figures~\ref{fig:scalingTimesteps} and~\ref{fig:scalingBatch} demonstrate that \sys's runtime scales linearly with increasing $T$ or $b$ due to the sequential operations over the timesteps or the linear increase in the number of ciphertexts processed. Finally, Figure~\ref{fig:scalingN} shows that \sys's local computation time remains constant with increasing $N$, as the data holders compute local operations at the same time whereas the communication overhead increases linearly, as more data holders are involved in the interactive protocols.

\subsection{Runtime Performance}\label{sec:runtimePerformance}
Table~\ref{table:runtime} displays \sys's training and prediction runtimes (for the original and optimized implementations, including communication) for various many-to-one RNN architectures (Elman) with $N{=}10$ data holders. The Training runtime is for one global iteration (including the communication for $\textsf{CBootstrap}(\cdot)$ and Aggregate Phase). The Prediction runtime is for oblivious predictions (both data and model are encrypted) on a batch of $256$ samples (including the communication for $\textsf{CKeySwitch}(\cdot)$ that changes the key of the prediction result to the querier's public key). The results of this table can be used along with the scalability results (Section~\ref{sec:scalability}) to estimate \sys's total training time for various RNNs. For instance, \sys can process 256K samples split among $10$ data holders with an RNN with $b{=}256, d{=}4, h{=}32, T{=}4$ (first table row), over $100$ global iterations, in $\sim$1.5 ($\sim$2) hours with its optimized (non-optimized, resp.) implementation. Moreover, as the Prediction runtime is calculated for 256 samples, \sys yields a prediction throughput of $\sim$14.43 ($\sim$9.31) samples per second with its optimized (non-optimized, resp.) implementation, for the same RNN structure. 

\begin{table*}[t]
\centering
\small
\renewcommand{\arraystretch}{0.9}
\begin{tabular}{cccccccccc}
\toprule
\multicolumn{1}{c}{ Dimension} &  \multicolumn{3}{c}{Jiang et al.~\cite{Jiang2018}} &  \multicolumn{3}{c}{\poseidon~\cite{poseidon}} & \multicolumn{3}{c}{Multi-dimensional ($\delta{=}32$)} \\ 
\midrule
$n\times n$  & Total | Amortized (s) & $\#M$  & $\log \mathcal{N}$ & Total | Amortized (s) & $\#M$  & $\log \mathcal{N}$ & Total | Amortized (s) & $\#M$  & $\log \mathcal{N}$ \\ \midrule
$32\times32$   & 0.43 | 0.05 & 8 & 14 & 6.14 | 6.14 & 1 & 14  & 0.43 | 0.05 & 8 & 14 \\ 
$64\times64$   & 0.82 | 0.41 & 2 & 14  & 14.02 | 14.02  & 1 & 14 & 1.70 | 0.21 & 8 & 14 \\ 
$128\times128$   & 2.86 | 2.86 & 1 & 15 & 69.25 | 69.25 & 1 & 15 & 7.16 | 0.89 & 8 & 14 \\ 
$256\times256$   & NA & 1 & 17 & NA & 1 & 17 & 30.6 | 3.82 & 8 & 14 \\ 
$512\times512$   & NA & 1 & 19 & NA & 1 & 17 & 79.17 | 9.89 & 8 & 14 \\ \bottomrule
\end{tabular}
\captionsetup{width=\linewidth}
\caption{Comparison between our multi-dimensional packing with Jiang et al.'s~\cite{Jiang2018} and Sav et al.'s (\poseidon)~\cite{poseidon} packing approaches.
We report the total and amortized runtime per matrix for the multiplication of $\#M$ matrices of size $n\times n$. $\log \mathcal{N}$ is the ring degree, and NA indicates that the memory was insufficient to carry out the evaluation.}
\label{table:matrixMultComparison}
\end{table*}

\descr{Effect of RNN Structures.}
 We evaluate \sys's performance for different RNN structures including many-to-one and many-to-many. Table~\ref{table:microLocal} depicts the results of microbenchmarks for the forward pass (FP) and backward pass (BP) \textit{per timestep} without bootstrapping and \textit{without optimization} for different RNN structures. FP-$t_o$ and BP-$t_o$ stand for the forward pass of the output-stage timestep and the backward pass of the output-stage timestep, respectively. The columns `FP Total' and `BP Total' show the total time required for the execution of the FP and BP, respectively. For example, the first row is a many-to-one RNN structure with $T{=}6$ and $\kappa{=}1$. This means that the FP is calculated for $6{-}1{=}5$ timesteps and FP-$t_o$ is calculated once. Similarly, for the second row with $T{=}6$ and $\kappa{=}2$, the FP is calculated for $6{-}2{=}4$ timesteps, whereas FP-$t_o$ is calculated for $2$. There is no FP and BP for the last row as $T{=}\kappa{=}6$. Overall, Table~\ref{table:microLocal} shows that the FP and BP calculations of the output-stage timestep (FP-$t_o$, BP-$t_o$) are more costly than the usual FP and BP timesteps ($\sim$1.7$\times$). This is due to the extra multiplication required in the FP for calculating the prediction (Line 5, Algorithm~\ref{algo:rnn}), and the extra subtraction and multiplication during the BP (Lines 11-12, Algorithm~\ref{algo:rnn}) for the output error calculation. Recall that Algorithm~\ref{algo:rnn} is presented for a many-to-many RNN structure with $\kappa{=}T$ outputs. The computations of Lines 5, 11, and 12, are executed $\kappa$ times for other structures with $\kappa{\neq}T$. Consequently, many-to-many RNN structures are more expensive, and the increase in the runtime depends on $\kappa$. We here note that a Jordan network requires the calculation of the output $y_t$ at every timestep (see Section~\ref{sec:background}). As a result, its computation time is very similar to a many-to-many Elman RNN with $T$ outputs and a modified matrix multiplication ($\mathbf{y}_{t-1}\times \mathbf{W}$). This results in roughly ${\sim}1.5\times$ slower execution per timestep compared to an Elman network. For instance, for the parameters of the first row in Table~\ref{table:microLocal}, a Jordan network requires $7.15$s and $10.02$s for the $FP$ and $BP$ without bootstrapping (similar to FP-$t_o$ and BP-$t_o$ of the line, resp.). Finally, note that the Jordan network requires one extra $\textsf{CBootstrap}(\cdot)$ operation during the backward pass, resulting in slightly higher communication than Elman networks.

\begin{table}[t]
\small
\centering
\setlength{\tabcolsep}{2.7pt}
\renewcommand{\arraystretch}{0.8}
\begin{tabular}{ccccccc}
\toprule
 & \multicolumn{6}{c}{Runtime (s)} \\ \cmidrule{2-7}
\begin{tabular}[c]{@{}c@{}}Structure\\ ($b$,$d$,$h$,$T$,$o$,$\kappa$)\end{tabular} & FP & FP-$t_o$ & BP & BP-$t_o$ & FP Total & BP Total \\
\midrule
(256, 64, 32, 6, 1, 1) & 4.36 & 7.44 & 7.60 & 11.55 & 29.24 & 49.55 \\
(256, 64, 32, 6, 1, 2) & 4.42 & 7.39 & 7.62 & 11.47 & 32.46 & 53.42 \\
(256, 64, 32, 6, 1, 4) & 4.32 & 7.47 & 7.55 & 11.59 & 38.52 & 61.46 \\
(256, 64, 32, 6, 1, 6) & - & 7.38 & - & 11.42 & 44.28 & 68.52\\ \bottomrule
\end{tabular}
\captionsetup{width=\linewidth}
\caption{Microbenchmarks for the forward pass (FP) and backward pass (BP) \textit{per timestep} \textit{without optimization}. $t_o$ represents the output-stage timestep.}
\label{table:microLocal}
\vspace{-2em}
\end{table}

\subsection{Comparison with Prior Work}\label{sec:microbenchmarks}
 We first compare \sys with prior work based on matrix multiplication as it is the dominant and expensive operation. Then, we compare \sys and the most similar system \poseidon~\cite{poseidon}.

\descr{Matrix Multiplication Benchmarks}. We compare our multi dimensional packing scheme with Jiang et al.'s~\cite{Jiang2018} and Sav et al.'s~\cite{poseidon} packing approaches in Table~\ref{table:matrixMultComparison} (timings) and Table~\ref{table:matrixMultScaling} (memory --  Appendix~\ref{sec:memory}). We report the total and amortized time (throughput) for $\#M$ multiplications of $n \times n$-sized matrices in parallel. 

We observe that our multi-dimensional packing approach and Jiang et al.'s one perform identically when $n{=}\delta$; since Jiang et al.'s approach is a special case of our multi-dimensional packing with only one ciphertext. For $n{>}\delta$ the multi-dimensional packing requires more time to complete the multiplication but achieves a better amortized time than Jiang et al's. Due to memory limitations, we were not able to benchmark Jiang et al.'s approach for $n{=}256$ and $n{=}512$ (see Table~\ref{table:matrixMultScaling}, Appendix~\ref{sec:memory}). The memory overhead arises from the ring dimension ($\mathcal{N}$), the number of ciphertexts, and the size of the evaluation keys. The first row of Table~\ref{table:matrixMultScaling} shows that, with Jiang et al.'s method, $\mathcal{N}$ scales quadratically with $n$ while ours scale similarly with $\delta$ (or with the closest power of two above $n^2$ or $\delta^2$, respectively). Thus, for larger matrices (e.g., when multiplying two $128\times 128$ matrices), Jiang et al.'s approach requires increasing $\mathcal{N}$ (as the matrix does not fit into one ciphertext), whereas our approach increases only the number of ciphertexts for a fixed $\delta$ (second row of Table~\ref{table:matrixMultComparison}). However, their approach also significantly impacts the size of the evaluation keys. Indeed, as in their approach $\mathcal{N}$ scales with $\mathcal{O}(n^2)$, increasing $n$ has the side effect of increasing the evaluation keys' size (third row of Table~\ref{table:matrixMultComparison}). Recall that Jiang et al.'s method also requires $\mathcal{O}(n)$ evaluation keys to perform the multiplication, thus resulting in a $\mathcal{O}(n^{3})$ memory complexity. In contrast, our approach provides a better memory complexity, with $\mathcal{O}(\delta^{3})$ that is $\mathcal{O}(1)$ for a fixed $\delta$. In conclusion, our packing is at least equivalent to Jiang et al.'s~\cite{Jiang2018} and can provide a better throughput for larger batch sizes or hidden dimensions such that the inputs/weights do not fit into one ciphertext. It also uses a constant memory for the evaluation keys, regardless of $n$, thus enabling the multiplication of large matrices, without the need for larger cryptographic parameters or the re-generation of the evaluation keys.

Sav et al.~\cite{poseidon} propose a system, \poseidon, for training feed-forward neural networks under MHE using a packing scheme that optimizes the learning over one sample (instead of a mini-batch) with vector-matrix operations. Thus, with their scheme, a matrix multiplication between two $n \times n$ matrices requires $n$ vector-matrix multiplications using their one-cipher packing approach (yielding same total and amortized time, see Table~\ref{table:matrixMultComparison}). With their approach a vector-matrix multiplication requires $\mathcal{O}(\log{n})$ inner sum rotations, and repeating this $n$ times to achieve a matrix multiplication yields a complexity of $\mathcal{O}(n \log{n})$ for a fixed $\mathcal{N}$. For example, for the multiplication of two $32 \times 32$ matrices, our multi-dimensional packing is $\sim$14$\times$ and $\sim$122$\times$ faster than \poseidon's packing approach for total and amortized times, respectively. The memory complexity of their approach also depends on $n$ for configuring $\mathcal{N}$, hence affecting the size of the evaluation keys (see Table~\ref{table:matrixMultScaling}, Appendix~\ref{sec:memory}), whereas our packing decouples the memory complexity from $n$.

Finally, we discuss another packing scheme similar to ours. 
Aharoni et al., improving on earlier work~\cite{aharoni2020}, propose a packing scheme that leverages on the CKKS complex domain to enable dense packing~\cite{aharoni2022}. Their approach is based on data structures that pack tensors, e.g., matrices or hypercubes, in fixed sized chunks, for a natural use of the SIMD capabilities. We observe that our multi-dimensional packing differs in multiple ways. First, their packing density is less optimal in terms of slot-usage as a ciphertext multiplication requires inner sums to compute the product between tensors. Consequently, the result will be sparsely packed compared to our technique that fully-utilizes the ciphertext slots. Secondly, their approach is not format-preserving thus requires pre- or post-processing, whereas our approach supports chaining multiplications without additional processing. Thus, even though their approach consumes two levels less than Jiang et al.~\cite{Jiang2018} (on which our packing is based) for the case of only one multiplication, the complexity of their scheme remains asymptotically equivalent. Finally, our approach is simpler to instantiate as its complexity depends only on $\delta$, whereas Aharoni et al.'s approach requires the configuration of many parameters for applications that require sequential operations, e.g., the training of RNNs. As their work does not provide an open source implementation, an experimental comparison is not possible.

\descr{Comparison with \poseidon~\cite{poseidon}.} We first note that a direct comparison between \sys and \poseidon is non-trivial as the latter does not support RNN training and important mechanisms such as gradient clipping. Moreover, \poseidon's packing scheme cannot handle RNN mini-batch training without introducing several changes, e.g., an explicit transpose operation required when the input batch is packed in one ciphertext, in its workflow. Yet, we estimate the runtime of \poseidon based on the matrix multiplication microbenchmarks (discussed earlier) and a naive-transpose operation. To process a batch of $b$ samples, \poseidon's one-sample processing approach requires the repetition of the forward and backward pass $b$ times, hence, it significantly increases computational complexity due to the large number of homomorphic rotations and vector-matrix multiplications that are required for the FP and BP of the various timesteps. Moreover, to support mini-batch training, \poseidon requires packing each sample to a different ciphertext, which significantly increases the communication overhead due to the required bootstrapping operations. For example, for a batch size of $b{=}256$ and $T{=}4$ \poseidon requires 1,024 $\textsf{CBootstrap}(\cdot)$ operations compared to 12 for \sys. Overall, we estimate that training an RNN with $d{=}32, h{=}32, b{=}256$, and $T{=}4$ using \poseidon would be $\sim$25-32$\times$ slower compared to \sys and its optimized version.

\section{Conclusion}

In this work, we presented \sys, a novel system that enables privacy-preserving training of and predictions on RNNs in a cross-silo federated learning setting. Building on multi-party homomorphic encryption (MHE), \sys preserves the confidentiality of the training data, the model, and the prediction data, under a passive adversary model with collusions of up to $N-1$ parties. By leveraging on a novel multi-dimensional packing scheme and polynomial approximations for clipping and activation functions, \sys enables efficient mini-batch training and addresses the problem of exploding/vanishing gradients that is inherent in RNNs. \sys scales sub-linearly with the number of features and the number of data holders, linearly with the number of timesteps and the batch size, and its accuracy is on par with non-secure centralized and decentralized solutions. To the best of our knowledge, \sys is the first system providing the building blocks for federated training of RNNs and its variants under encryption. As future work, we plan to evaluate \sys on more complex RNN architectures.


\bibliographystyle{abbrv}
\bibliography{bibfile}

\newcounter{subsubsubsection}
\appendices

\section{\sys Extensions} \label{sec:extensions}

\descr{Network Topology.} As discussed in Section~\ref{sec:design}, \sys relies on a tree-structure for efficiently aggregating the data holders' local models in the FL protocol. Yet, it is worth mentioning that \sys's underlying operations are decoupled from the network topology as each party performs local RNN training under encryption. For instance, switching to the traditional FL star topology (with a central aggregation server) would require the server to add the $N$ models received by the parties. Furthermore, the communication complexity of the collective bootstrapping ($BS_c$) and the FL workflow ($FL_c$) (see Appendix~\ref{sec:complexityDetailed}) would be affected proportionally to the number of links between the server and the clients, i.e., with a star topology the $N-1$ terms of both $BS_c$ and $FL_c$ (see Appendix~\ref{sec:complexityDetailed}) would be $N$.

\descr{Availability, Dropouts, and Asynchronous Learning.} In this work, we assume that all data holders are available (online) during \sys's training and prediction workflows. As we rely on a cross-silo federated learning setting, this assumption is realistic, yet, prohibitive in terms of dropouts as the collective cryptographic operations, e.g., $\textsf{CBootstrap}(\cdot)$ or $\textsf{CKeySwitch}(\cdot)$, require collaboration among all the data holders. Thus, to support asynchronous federated learning (driven by a time-threshold or by relying on a subset of online data holders) or dropouts, \sys can be deployed with a threshold multi-party homomorphic encryption scheme which enables a subset of $t$-out-of-$N$ data holders to perform the collective operations~\cite{Mouchet2022}. However, this cryptographic scheme would introduce a relaxation in \sys's threat model (allowing collusions of up to $t-1$ data holders instead of $N-1$).


\descr{Active Adversaries During Training.} As discussed in Section~\ref{sec:design}, \sys assumes a passive-adversary threat model that tolerates collusions between up to $N-1$ parties. Extending \sys to active adversaries requires primitives that enable the verification of the tree-based aggregation and the local gradient descent operations performed by each data holder. Such primitives can be implemented with verifiable computation techniques, e.g., secure multi-party computation~\cite{stevens2022efficient} or zero-knowledge proofs~\cite{zheng2019helen}, however, with a significant computational cost. Moreover, another extension required to tolerate active adversaries during training is to verify the correctness of the data holders' inputs, e.g., using statistical tests~\cite{chen2021holmes} or proofs of authenticity on encrypted data~\cite{chatel2021privacy}, as well as their consistency, e.g., using cryptographic commitments~\cite{zheng2019helen}. While such techniques can reduce the risk of poisoning attacks in federated learning~\cite{tolpegin2020data}, their elimination remains an open research problem. Currently proposed defenses rely on gradient inspection~\cite{Li2020,Zhang2022,Tolpegin2020,Yang2021,Alistarh2018}, thus, raising new challenges for encrypted FL pipelines.


\descr{Attacks Against PaaS.} As discussed in Section~\ref{sec:design}, by retaining any information exchanged among the data holders and the final model encrypted, \sys mitigates inference attacks that target the gradients/model updates exchanged during the FL training process~\cite{Nasr2019, Melis2019, Hitaj2017, Zhu2019}. Nonetheless, inference attacks, e.g., model stealing~\cite{tramer2016stealing} or membership inference~\cite{shokri2017membership}, are also possible by exploiting the outputs of the PaaS. This is particularly the case for time-series applications (e.g., auto-regressive ones) where predictions (or queries) might be performed continuously over time. \sys can minimize such leakage by feeding encrypted predictions back to the RNN for further oblivious processing and decrypt only an aggregated result (e.g., revealing only the time-series trend over a coarse period instead of a more fine-grained one). Additionally, such inference attacks against PaaS can be mitigated with mechanisms orthogonal to \sys, e.g., by limiting the number of predictions a querier can request and/or by integrating a mechanism that perturbs the prediction outputs to obtain differential privacy guarantees.

\section{Security Analysis}\label{sec:security}
We demonstrate that \sys fulfills the data and model confidentiality objectives (Section~\ref{sec:problemStatement}) by arguing that a computationally-bounded adversary that controls up to $N-1$ data holders (i.e., yielding a collusion among $N-1$ data holders) cannot infer any information about the honest data holder's data or the trained model. We first note that CKKS scheme is IND-CPA secure and its semantic security is based on the hardness of the decisional RLWE problem~\cite{cheon2017homomorphic,Lyubashevsky2010,Lindner2011}. We rely on the proofs by Mouchet et al.~\cite{mouchet2019distributedbfv}, which show that the MHE cryptographic protocols, i.e., $\textsf{CKeyGen}(\cdot)$ and $\textsf{CKeySwitch}(\cdot)$, are secure in a passive-adversary model with up to $N-1$ collusions as long as the underlying RLWE problem is hard. While their proofs are constructed for the BFV scheme, they generalize to CKKS, since the two schemes rely on the same computational assumptions. The security of the collective bootstrapping operation, i.e., $\textsf{CBootstrap}(\cdot)$, is proven analogously in earlier works employing CKKS-based MHE~\cite{spindle,poseidon}.

Assume that any ciphertext communicated with \sys is generated using the CKKS cryptosystem parameterized to ensure a post-quantum security level of $\lambda$. We rely on the real/ideal world simulation paradigm~\cite{lindell2017simulate} and consider a real-world simulator $\mathcal{S}$ that simulates a computationally-bounded adversary corrupting $N-1$ data holders. During \sys’s training protocol, the model parameters that are exchanged among data holders at the Aggregate and Model Update Phases are encrypted with the collective public key, while all other phases rely on the collective MHE operations that are proven to be CPA-secure. The result of any computation performed during the Local Computation Phase, following Lemmas 2, 3, and 4 of~\cite{cheon2017homomorphic}, is a valid encryption under CKKS, thus, achieves a security level of $\lambda$. Furthermore, the decryption of any ciphertext requires collaboration among \textit{all} the data holders that participated in $\textsf{CKeyGen}(\cdot)$. To avoid leakage due to two consecutive broadcasts, we rely on an existing countermeasure~\cite{mouchet2019distributedbfv} that re-randomizes any ciphertext before communicating it to the network. As a result, the sequential composition of all cryptographic computations during the training phase of \sys remains simulatable by $\mathcal{S}$. Similarly, during \sys's prediction phase, the data of the querier is encrypted with the collective public key and the prediction result is re-encrypted with the querier's public key; this is the only entity that can decrypt thanks to the simulatable $\textsf{CKeySwitch}(\cdot)$ functionality. Hence, $\mathcal{S}$ can simulate all of the encryptions communicated during \sys's training and prediction phase by generating random ciphertexts with equivalent security parameters; the real outputs cannot be distinguished from the ideal ones. Consequently, \sys protects the confidentiality of the honest data holder's data and the model (following analogous arguments).

\section{Parameter Selection}\label{sec:paramSelection}
We now devise guidelines for choosing the parameters that play a vital role for the efficient RNN training execution in \sys. In particular, we discuss  the cryptographic parameters and the sub-matrix dimension $\delta=n/\ell$ (see Section~\ref{sec:packing}) that is crucial for the efficiency of the multi-dimensional matrix multiplication (Algorithm~\ref{algo:matrixmultiplicationCipher}). RNN-related parameters, e.g., the dimension of the hidden matrix, depend on the learning task and the characteristics of the input data, thus, their configuration is out-of-the-scope of this work.

\descr{Cryptographic Parameters.} These are linked to the depth of the circuit under evaluation and the desired security level. The circuit depth highly depends on the number of hidden layers and the approximation degree ($\mathfrak{p}$) of the activation and clipping functions. A higher circuit depth implies a higher number of initial level $\mathcal{L}$ and $Q$, for reducing the number of $\textsf{CBootstrap}(\cdot)$ yielding a requirement for bigger cryptographic parameters. To configure the security level, \sys follows the guidelines of the homomorphic encryption standardization whitepaper for choosing the cyclotomic ring size $\mathcal{N}$ (given the ciphertext modulus $Q$)~\cite{HEStandardWeb}. Configuring the cryptographic parameters for the training of neural networks in general is a non-trivial task due to the high number of other parameters that affect this choice, e.g., the number of layers, the polynomial approximation degree, the number of data holders, etc. We refer the reader to~\cite{poseidon} for additional details on how to configure the cryptographic parameters based on other neural network learning parameters and to~\cite{mouchet2019distributedbfv} for an overview of these parameters.

\descr{Choosing $\delta$.} Recall that the complexity of the matrix multiplication described in Algorithm~\ref{algo:matrixmultiplicationCipher} amortizes to $\mathcal{O}(\delta)$. In the extreme case where $\delta=1$, there are no costly ciphertext operations, except for arithmetic multiplications and additions (no linear transformations are needed). Thus, choosing $\delta=1$ minimizes the amortized computational complexity and supports any dimension for the input matrices. However, the memory complexity (in the number of ciphertexts) becomes the number of elements in the input matrices. Since batch sizes are usually in the range $\{64, 128, 256\}$, it is unlikely that all the $\mathcal{N}/2$ ciphertext slots will be used when setting $\delta=1$. In fact, since the number of slots is usually in the range of $\{2^{13}, 2^{14}, 2^{15}\}$, this approach would lead to a very low packing density and inefficient memory usage. Instead, we choose a value for $\delta$ that provides an acceptable trade-off between computational cost and memory complexity. 
To maximize the packing density, the batch size should be $b = \frac{\mathcal{N}}{2*\delta}$, thus we can derive that $\delta = \frac{\mathcal{N}}{2*b}$. Since the batch size can be small due the learning task, we introduce the ciphertext-utilization parameter $\alpha$ that indicates the fraction of the utilized ciphertext slots, to relax the utilization assumption (e.g., $\alpha=1$ and $\alpha=0.5$ indicate full and half utilization, respectively). Given that the batch size cannot be more than $maxB$ and that the ciphertext utilization is at least $\alpha$ we have:
\[
  \frac{\mathcal{N}*\alpha}{2*\delta}\le  b \le maxB
\]
which translates to:
\[ \delta \ge \frac{\mathcal{N}*\alpha}{2*maxB}.\]\\
For example, in our experiments (see Section~\ref{sec:evaluation}), we choose $\mathcal{N}=2^{14}$. Thus, for a full ciphertext utilization ($\alpha=1$) at $b=256$, we have: 
\[\delta \ge \frac{2^{14}}{2*256} = 32.\]\\
For a ciphertext utilization of $\alpha=0.5$, the batch size can be as small as $b=128$ at the same $\delta=32$, following a similar calculation.

\begin{figure*}[t]
    \centering
    \tiny
    \begin{subfigure}[b]{0.40\textwidth}
        \centering
        \includegraphics[width=0.8\textwidth]{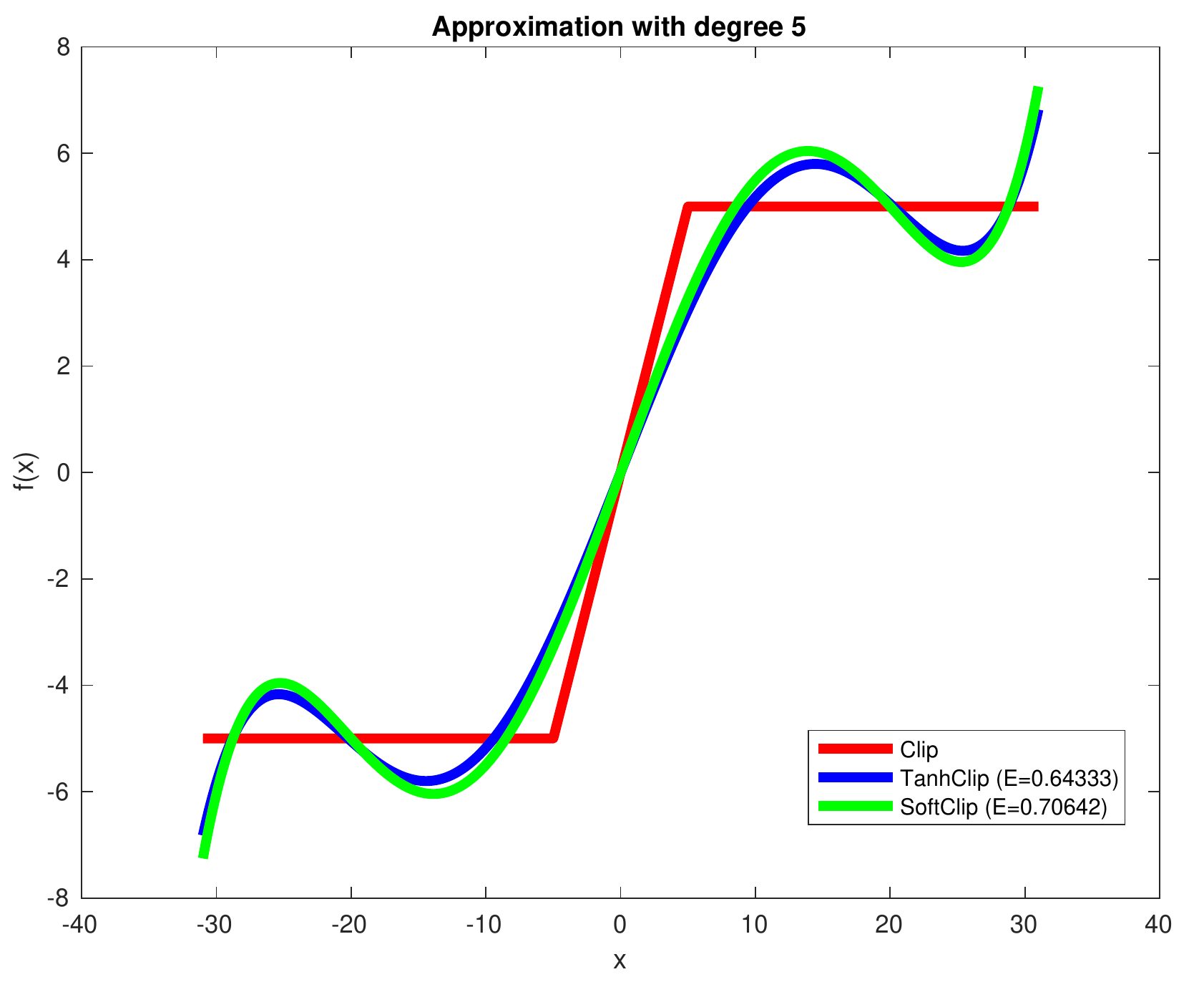}
        \caption{Approximation curves with $\mathfrak{p}=5$}  
        \label{fig:deg5}
    \end{subfigure}
    \begin{subfigure}[b]{0.40\textwidth}  
        \centering 
        \includegraphics[width=0.8\textwidth]{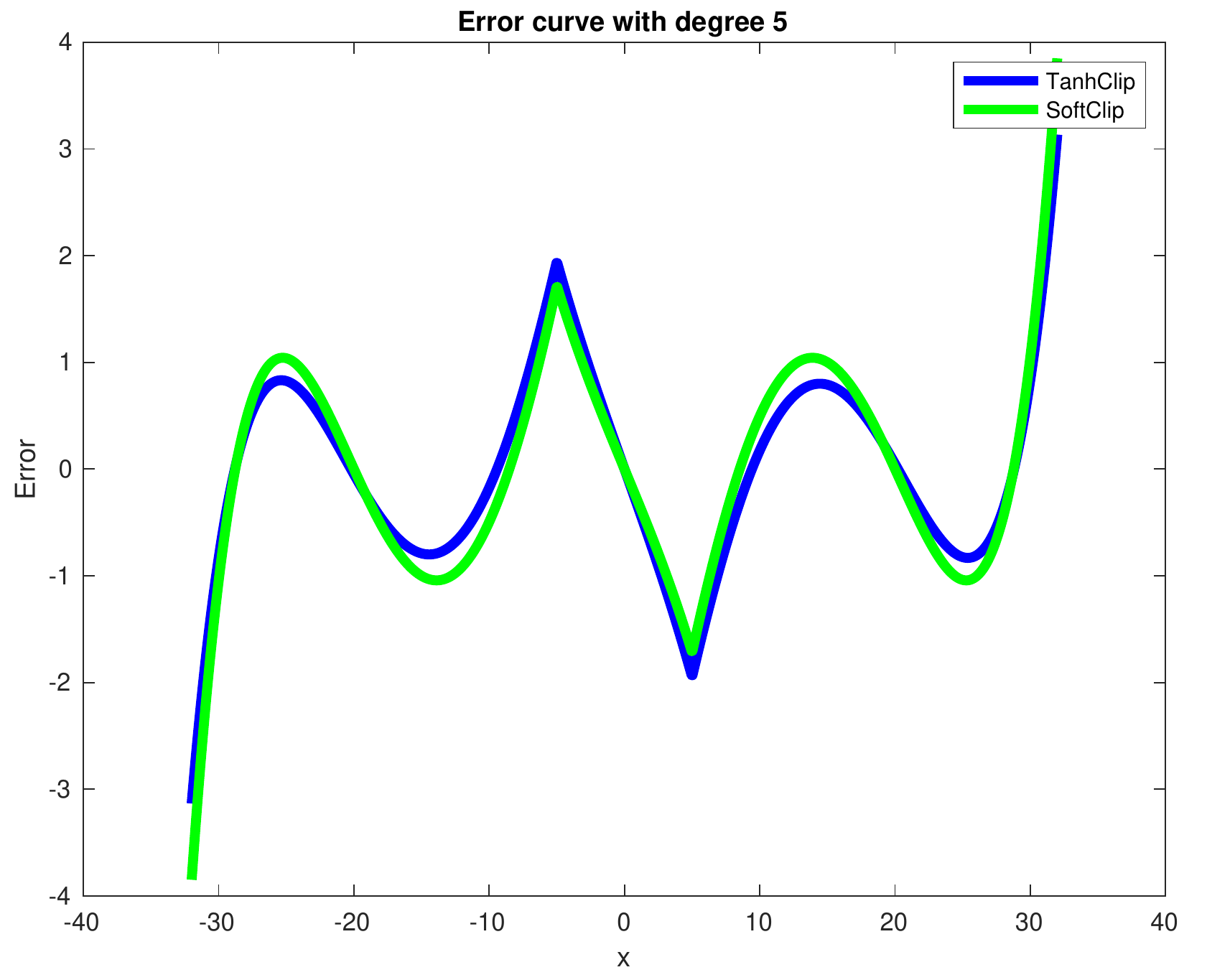}
        \caption{Error curves with $\mathfrak{p}=5$}  
        \label{fig:deg5Err}
    \end{subfigure}
    \vskip\baselineskip
    \begin{subfigure}[b]{0.40\textwidth}   
        \centering 
        \includegraphics[width=0.8\textwidth]{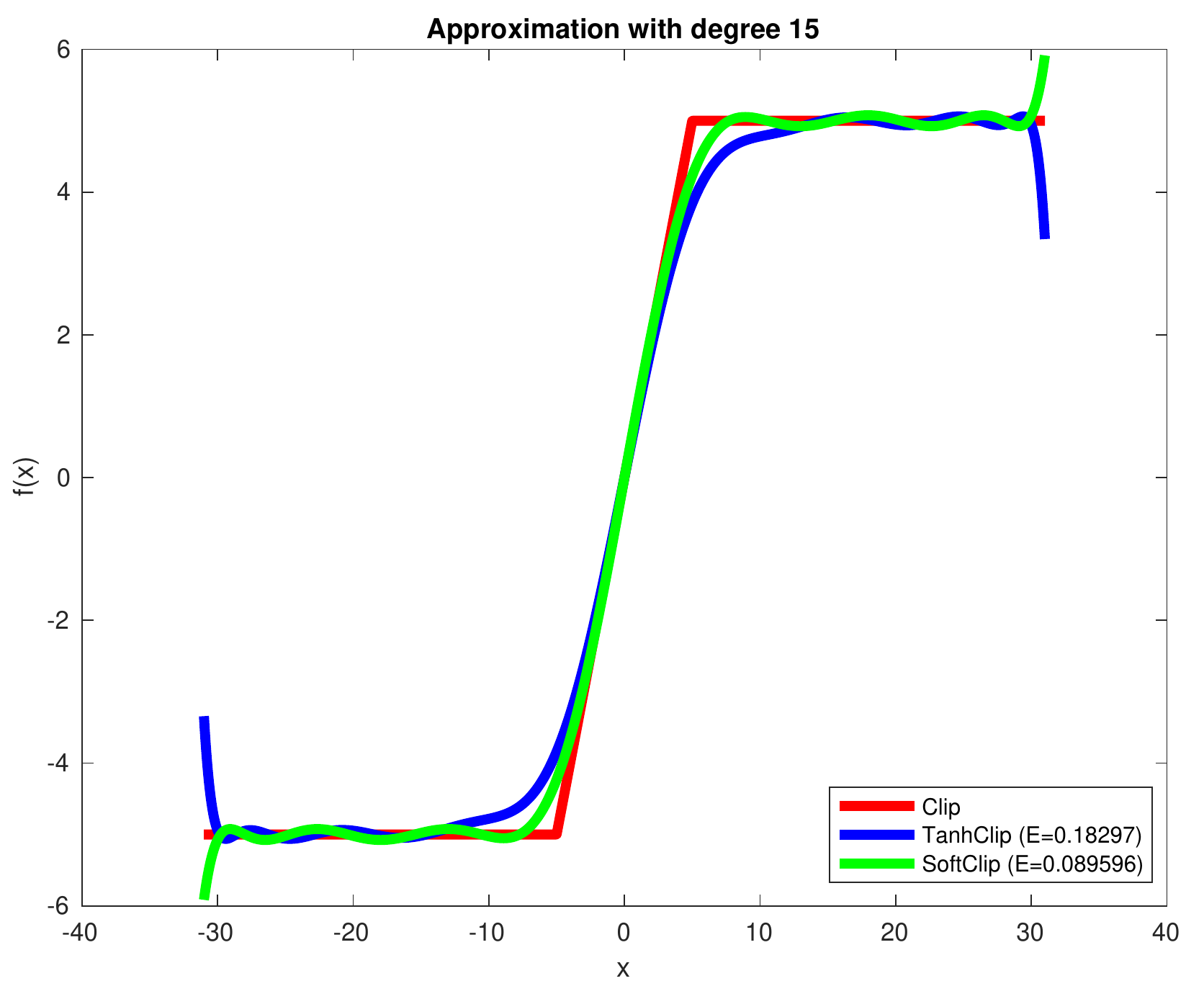}
        \caption{Approximation curves with $\mathfrak{p}=15$} 
        \label{fig:deg15}
    \end{subfigure}
    \begin{subfigure}[b]{0.40\textwidth}   
        \centering 
        \includegraphics[width=0.8\textwidth]{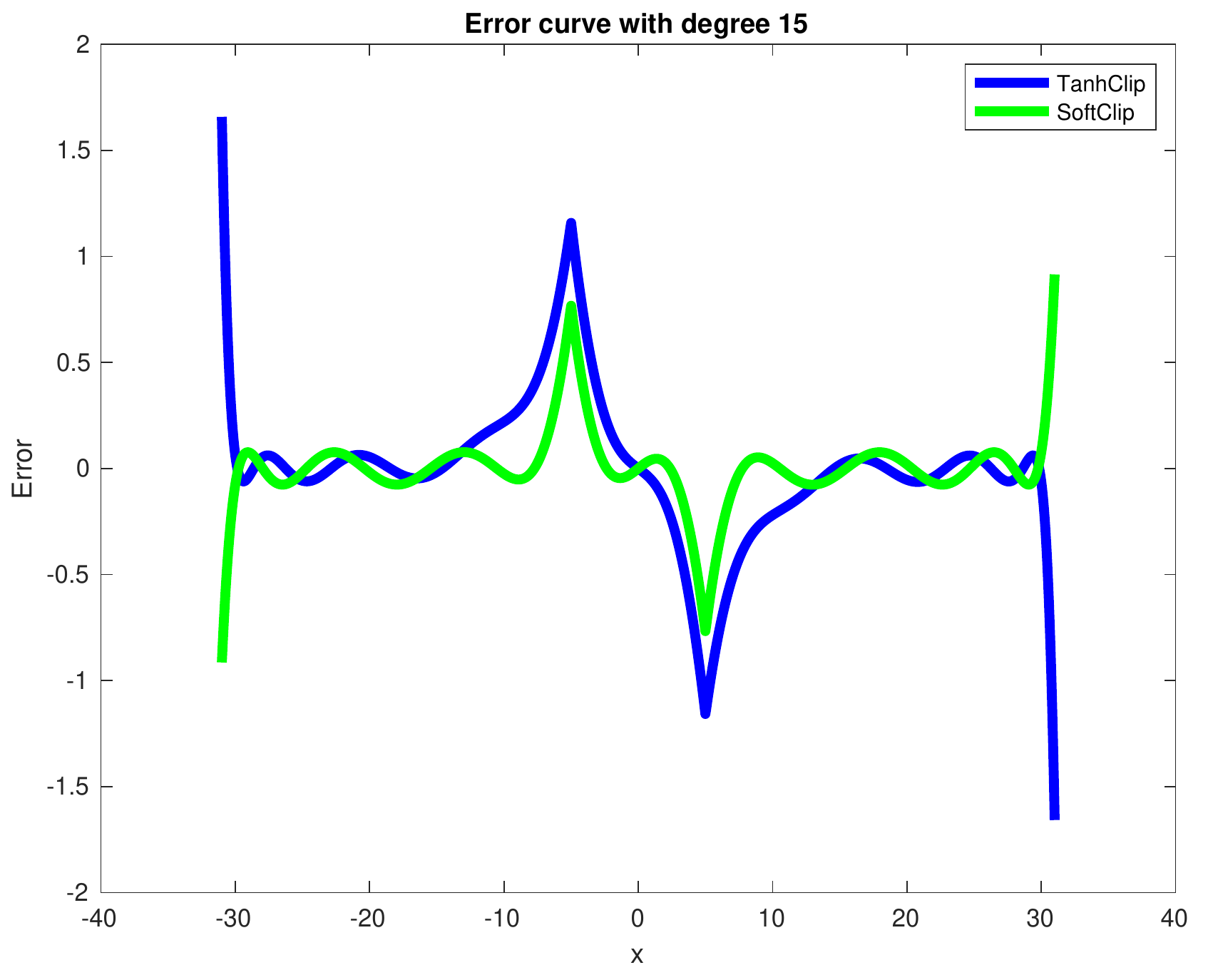}
        \caption{Error curves with $\mathfrak{p}=15$}  
       \label{fig:deg15Err}
    \end{subfigure}
    \caption{Approximation and error curves of $\textsf{TanhClip}(\cdot)$ and $\textsf{SoftClip}(\cdot)$ compared to baseline clipping function $\textsf{Clip}(\cdot)$ with degrees $\mathfrak{p}=5$ and $\mathfrak{p}=15$ in the interval $[-30,30]$ for a clipping threshold $|m|=5$. Error curves are plotted in the range of $[-32,32]$. $E$ in the legends of the  plots indicates the average absolute error per approximation in the interval $[-30,30]$.}
    \label{fig:approx}
\end{figure*}

\section{Approximation of the proposed clipping functions}\label{sec:approxCurve}
Figure~\ref{fig:approx} displays the approximation and error curves of $\textsf{TanhClip}(\cdot)$ and $\textsf{SoftClip}(\cdot)$ (compared to the baseline clipping function ($\textsf{Clip}(\cdot)$)) by employing the Minimax method with degrees $\mathfrak{p}=5$ and $\mathfrak{p}=15$ in the approximation interval $[-30,30]$ and for a clipping threshold of $|m|=5$. Error curves are plotted in the range of $[-32,32]$ to observe the behavior of the approximations out of the approximation interval limits. $E$ in the plot legends indicates the average absolute error of each function in the approximation interval $[-30,30]$. We observe that both approximations yield similar shapes and their error oscillation is alike. Yet, for smaller degrees, e.g., $\mathfrak{p}=5$, the average error $E$ of $\textsf{TanhClip}(\cdot)$ is slightly smaller than $\textsf{SoftClip}(\cdot)$ ($0.643$ vs.$0.706$). For higher degrees ($\mathfrak{p}=15$), however, $\textsf{SoftClip}(\cdot)$ achieves a better approximation (i.e., $E=0.089$ vs. $E=0.183$ for $\textsf{TanhClip}(\cdot)$). Moreover, the divergence of $\textsf{SoftClip}(\cdot)$ near the limits of the approximation interval is slower than $\textsf{TanhClip}(\cdot)$ (see Figure~\ref{fig:deg15}); this is a desired characteristic for any approximation. As a consequence, one should choose the approximation method based on the characteristics of the dataset, the desired degree (or precision) required for the polynomial approximation, and the aforementioned features of these approximations.



\section{Local Computation Phase for GRUs}\label{sec:gru_algo}
Algorithm~\ref{algorithm:gru} summarizes \sys's local computation phase, i.e., the forward and backward pass, for training a GRU.
\begin{algorithm}
  \caption{Local Computation Phase (GRU)}\label{algorithm:gru}
  \begin{algorithmic}[1]
\Require  $\mathbf{U}_z,\mathbf{U}_r,\mathbf{U}_n, \mathbf{W}_z,\mathbf{W}_r,\mathbf{W}_n, \mathbf{V}, \mathbf{b}_y, \mathbf{b}_z, \mathbf{b}_r, \mathbf{b}_n, \mathbf{h}_{prev},$ \Statex $X = (x_1,x_2,\cdots,x_t), Y = (y_1,y_2,\cdots,y_t)$ 
\Ensure $\nabla\mathbf{U}_z,\nabla\mathbf{U}_r,\nabla\mathbf{U}_n, \nabla\mathbf{W}_z,\nabla\mathbf{W}_r,\nabla\mathbf{W}_n, \nabla\mathbf{V}, \nabla\mathbf{b}_y, \nabla\mathbf{b}_z, \nabla\mathbf{b}_r, \nabla\mathbf{b}_n$
  \State $\mathbf{h}_{0} \gets \mathbf{h}_{prev}$
    \For{$t \gets 1 : T$} \Comment{Forward Pass}
            \State $\mathbf{z}_{raw} \gets \mathbf{h}_{t-1} \times \mathbf{W}_z + x_t \times \mathbf{U}_z + \mathbf{b}_z$ \Comment{Z-Gate}
        \State $\mathbf{z}_t \gets \varrho(\mathbf{z}_{raw})$
        \State $\mathbf{r}_{raw} \gets \mathbf{h}_{t-1} \times \mathbf{W}_r + x_t \times \mathbf{U}_r + \mathbf{b}_r$ \Comment{R-Gate}
        \State $\mathbf{r}_t \gets \varrho(\mathbf{r}_{raw})$ 
        \State $\mathbf{n}_{raw} \gets (\mathbf{r}_t\odot \mathbf{h}_{t-1}) \times \mathbf{W}_n + x_t \times \mathbf{U}_n + \mathbf{b}_n$ \Comment{N-Gate}
        \State $\mathbf{n}_t \gets \varphi(\mathbf{n}_{raw})$
        \State $\mathbf{h}_t \gets \mathbf{z}_t \odot \mathbf{h}_{t-1} + (1-\mathbf{z}_t)\odot \mathbf{n}_t$ \Comment{Hidden State}
        \State $\mathbf{p}_t \gets \mathbf{h}_t \times \mathbf{V} + \mathbf{b}_y$ \Comment{Output Gate}
    \EndFor
    \State $\mathbf{dh}_{nxt} = \mathbf{0}$ \Comment{$\frac{\partial loss[T+1:\infty]}{\partial \mathbf{h}_t}$}
    \State $\nabla\mathbf{U}_z,\nabla\mathbf{U}_r,\nabla\mathbf{U}_n, \nabla\mathbf{W}_z,\nabla\mathbf{W}_r,\nabla\mathbf{W}_n, \nabla\mathbf{V}, $
    \Statex $\nabla\mathbf{b}_y, \nabla\mathbf{b}_z, \nabla\mathbf{b}_r, \nabla\mathbf{b}_n = \mathbf{0}$
    \For{$t \gets T : 1$} \Comment{Backward Pass}
        \State $\mathbf{dy} \gets \mathbf{p}_t - y_t$ 
        \State $\mathbf{dh} \gets \mathbf{dy} \times \mathbf{V}^T  + \mathbf{dh}_{nxt}$ 
        \State $\mathbf{dn} \gets (1-\mathbf{z}_t)\odot \mathbf{dh}$ 
        \State $\mathbf{dn}_{raw} \gets \varphi'(\mathbf{n}_t) \odot \mathbf{dn} = (1-n^2_t) \odot \mathbf{dn}$ 
        \State $\mathbf{dr} \gets ( \mathbf{dn}_{raw} \times \mathbf{W}_n^T) \odot \mathbf{h}_{t-1}$ 
        \State $\mathbf{dr}_{raw} \gets \varrho'(\mathbf{r}_t) \odot \mathbf{dr} = \mathbf{r}_t\odot(1-\mathbf{r}_t) \odot \mathbf{dr}$ 
        \State $\mathbf{dz} \gets (\mathbf{h}_{t-1}-\mathbf{n}_t) \odot \mathbf{dh}$ 
        \State $\mathbf{dz}_{raw} \gets \varrho'(\mathbf{z}_t) \odot \mathbf{dz} = \mathbf{z}_t\odot(1-\mathbf{z}_t) \odot \mathbf{dz}$ 
        \State $\mathbf{dh}_h \gets \mathbf{dh} \otimes \mathbf{z}_t $ 
        \State $\mathbf{dh}_z \gets \mathbf{dz}_{raw} \times \mathbf{W}_z^T$ 
        \State $\mathbf{dh}_r \gets \mathbf{dr}_{raw} \times \mathbf{W}_r^T $ 
        \State $\mathbf{dh}_n \gets \mathbf{r}_t \otimes (\mathbf{dn}_{raw} \times \mathbf{W}_n^T)  $ 

        \State $\mathbf{dh}_{nxt} \gets \mathbf{dh}_z+\mathbf{dh}_h+\mathbf{dh}_n+\mathbf{dh}_r$ 
        
        \State $\nabla\mathbf{V} += \mathbf{h}_t \otimes \mathbf{dy}$ 
        \State $\nabla\mathbf{U}_z += x_t \otimes \mathbf{dz}_{raw}$ 
        \State $\nabla\mathbf{U}_r +=  x_t \otimes \mathbf{dr}_{raw} $ 
        \State $\nabla\mathbf{U}_n += x_t \otimes \mathbf{dn}_{raw}$ 

        \State $\nabla\mathbf{W}_z +=  \mathbf{h}_{t-1} \otimes \mathbf{dz}_{raw}$ 
        \State $\nabla\mathbf{W}_r +=  \mathbf{h}_{t-1} \otimes \mathbf{dr}_{raw}$ 
        \State $\nabla\mathbf{W}_n += \mathbf{r}_t \odot (\mathbf{h}_{t-1} \otimes \mathbf{dn}_{raw})$ 

        \State $\nabla\mathbf{b}_y += \mathbf{dy}$ 
        \State $\nabla\mathbf{b}_z += \mathbf{dz}_{raw}$ 
        \State $\nabla\mathbf{b}_r += \mathbf{dr}_{raw}$
        \State $\nabla\mathbf{b}_n += \mathbf{dn}_{raw}$ 

    \EndFor
\end{algorithmic}
\end{algorithm}
\section{Detailed Complexity Analysis}\label{sec:complexityDetailed}

Here we detail the analysis on \sys's memory cost per data holder, its communication and computational costs for an Elman network given in Section~\ref{sec:complexity}. We account for the Local Computation, Aggregate, and Model-Update Phases. We exclude the Setup Phase, i.e., the generation of the cryptographic keys ($pk$, $[ek]$, etc.) and their memory costs from our analysis; this is a one-time phase whose communication cost predominantly depends on the generation of the rotation and relinearization keys (see ~\cite{mouchet2019distributedbfv} for details).


Recall that the number of ciphertext slots is ${\mathcal{N}}/{2}$, the input size (number of features) is denoted as $d$, the batch size as $b$, the hidden dimension as $h$, the sub-matrix dimension as $\delta$, the output size as $o$, and the number of data holders, local iterations, global iterations, and timesteps as $N$, $l$, $e$, and $T$, respectively. The maximum size of a ciphertext is denoted by $s$. We denote the number of plaintexts used to pack a vector/matrix $a$ as $|a|$, and the number of ciphertexts with bold-face \bm{$a$}. To ease the presentation of our complexity analysis we use the following configuration as an example:

\descrit{Example Configuration:} Assume a ring size of $\mathcal{N}=2^{14}$ and ciphertexts with an initial level $\mathcal{L}=10$. Then, assume that $d=16$, $b=256$, $h=32$, $\delta=32$, $\kappa=1$ (many-to-one RNN), and $o=1$ (i.e., a regression task). Finally, assume a degree 7 ($\mathfrak{p}=7$) polynomial approximation of the activation and clipping functions (these are parameters primarily used in our experiments).

\descr{Memory Cost.}  We estimate the memory usage per data holder based on the number of input plaintexts, activation ciphertexts (e.g., $\mathbf{h}_t$ in Algorithm~\ref{algo:rnn}), weight ciphertexts (e.g., $\mathbf{U},\mathbf{V},\mathbf{W}$), and gradient ciphertexts (e.g., $\mathbf{\nabla U},\mathbf{\nabla W},\mathbf{\nabla V}$). We denote the number of plaintexts used to pack a vector/matrix $a$ as $|a|$, and the number of ciphertexts with bold-face \bm{$a$}. We first calculate the number of matrix rows/columns that each ciphertext can pack. Since we rely on row-based packing, the number of matrix rows per ciphertext is $\floor{\frac{\mathcal{N}}{2*\delta}}$ and the number of columns per ciphertext is $\delta$, for any matrix size. As we split the input batch over the third dimension (i.e., the size of a ciphertext), the number of sub-matrix rows per ciphertext is $\ceil{\frac{b}{(\mathcal{N}/2)/\delta}}= \ceil{\frac{2*\delta*b}{\mathcal{N}}}$ and the number of sub-matrix columns is $\ceil{\frac{c}{\delta}}$, for any column of size $c$. These allow us to derive the total number of plaintexts required for the input batch and \textbf{activation} ciphertexts as:
 
\[|x_t| = \ceil{\frac{2*\delta*b}{\mathcal{N}}}* \ceil{\frac{d}{\delta}} \text{\quad(input plaintexts)}\]
\[|\mathbf{z}_t| = |\mathbf{h}_t| =  \ceil{\frac{2*\delta*b}{\mathcal{N}}}* \ceil{\frac{h}{\delta}}\text{\quad(hidden ciphertexts)}\]
\[|\mathbf{p}_t| = \ceil{\frac{2*\delta*b}{\mathcal{N}}}* \ceil{\frac{o}{\delta}}\text{\quad(output ciphertexts)}\]

\noindent As we replicate the weight matrices for the sub-matrix operations (see Section~\ref{sec:packing}), for an $n\times m$ weight matrix, the number of sub-matrix rows and columns per weight ciphertext is $\ceil{\frac{n}{\delta}}$ and $\ceil{\frac{m}{\delta}}$, respectively. Therefore, the total number of ciphertexts for the \textbf{weight} matrices are:
\[|\mathbf{U}| = \ceil{\frac{d}{\delta}}*\ceil{\frac{h}{\delta}}\text{\quad(input-weight ciphertexts)}\]
\[|\mathbf{W}| = \ceil{\frac{h}{\delta}}*\ceil{\frac{h}{\delta}}\text{\quad(hidden-weight ciphertexts)}\]
\[|\mathbf{V}| = \ceil{\frac{h}{\delta}}*\ceil{\frac{o}{\delta}}\text{\quad(output-weight ciphertexts)}\]

\noindent Finally, the number of gradient ciphertexts per weight matrix is the same as the number of ciphertexts for the respective activation or weight matrix. The number of plaintexts and ciphertexts sets an upper bound to the memory cost per data holder, per timestep execution. Yet, not all computed values need to be stored during the training. For example, when the activation function is $\textsf{Tanh}$, $\mathbf{dh} \odot \varphi'(\mathbf{z}_t)$ (Line 13, Algorithm~\ref{algo:rnn}) is equivalent to $\mathbf{dh} \odot (1-\mathbf{h}^2_t) $ as the derivative of $\textsf{Tanh}(x)$ is $1-\textsf{Tanh}^2(x)$. Thus, $\mathbf{z}_t$ is computed on-the-fly in the forward pass, and then discarded as it is not used in the backward one. For the Example Configuration, \sys requires only one ciphertext for each activation, weight, and gradient matrix.


\descr{Communication Cost.} Recall that the communication among the data holders is triggered by: (a) the collective bootstrapping ($\textsf{CBootstrap}(\cdot)$) operation to refresh the ciphertexts, and (b) the federated learning (FL) workflow, where data holders aggregate their locally computed gradients (i.e., Aggregate Phase) before the \textit{aggregation server} updates the global model and broadcasts it (i.e., Model-Update Phase).

\descrit{Collective Bootstrapping.} The total number of bootstrapping operations depends on both the learning and the cryptographic parameters. For instance, the approximation degree of the activation or clipping functions ($\mathfrak{p}$) and the polynomial ring dimension ($\mathcal{N}$) have a direct effect on the depth of the circuit and thus, the number of bootstraps. We refer the reader to~\cite{poseidon} for the detailed estimation of the number of bootstraps depending on the learning and the cryptographic parameters. For the Example Configuration, \sys bootstraps $\mathbf{h}_t$ (Line 4) and $\mathbf{dz}$ (Line 13) in the forward pass of each local timestep iteration (Algorithm~\ref{algo:rnn}). After the execution of all timesteps, the intermediate term $\mathbf{I} = \mathbf{h}_t \otimes \mathbf{dy}$ (Line 15) in the backward pass is bootstrapped in the output stage, i.e., \sys executes one bootstrap for a one/many-to-one RNN structure and $\kappa$ bootstraps for one/many-to-many structures on $\mathbf{I}$ where $|\mathbf{I}|=|\mathbf{dz}|$. \sys also bootstraps the weight ciphertexts $\mathbf{U}$, $\mathbf{V}$, and $\mathbf{W}$, after the Model-Update Phase to refresh them before the next training round. Thus, the communication cost due to the collective bootstrapping operation ($BS_c$) is: 
\[BS_c < {e(|\mathbf{U}|+|\mathbf{W}|+|\mathbf{V}|+l|\mathbf{I}|\kappa+Tl(|\mathbf{ht}|+|\mathbf{dz}|))(N-1)s}.\]
Recall that $s$ sets an upper bound for the messages; the communication cost of the bootstrapping is lower in practise. For the Example Configuration $BS_c=(4+2T)(N-1)s$ per local iteration.

\descrit{FL Workflow.} During a global training iteration, the data holders collectively aggregate their locally computed gradients (i.e., $\nabla\mathbf{U}$, $\nabla\mathbf{W}$, $\nabla\mathbf{V}$, see Line 8, Algorithm~\ref{algo:fedavg}) and the aggregation server updates the global model and broadcasts its weights (i.e., $\mathbf{U}$, $\mathbf{W}$, $\mathbf{V}$, Line 4, Algorithm~\ref{algo:fedavg}). In \sys, data holders are organized in a tree-structured network topology, thus, they collectively aggregate their gradients by sending them to their parent in the tree. Then, the aggregation server broadcasts the updated weights down the tree. Given that this communication is incurred for every global iteration, the total communication cost ($FL_c$) is:
\[FL_c={(|\nabla\mathbf{U}|+|\nabla\mathbf{W}|+|\nabla\mathbf{V}|+|\mathbf{U}|+|\mathbf{W}|+|\mathbf{V}|)e(N-1)s}.\]

\descr{Computational Cost.} We remind that \sys's computational cost per data holder and per local iteration depends on the \textit{dominating terms}, i.e., the number of $\textsf{CBootstrap}(\cdot)$ operations, the degree $\mathfrak{p}$, and the cost of the linear transformations (i.e., $\sigma$, $\tau$, $\psi_i$ and $\phi_i$) for the matrix multiplication. Note that bootstrapping a ciphertext at a level $\mathcal{L}_{\bm{c}}$ (with an initial level $\mathcal{L}$) has a complexity of $C(BS)=\mathcal{N}\log_{2}(\mathcal{N})(\mathcal{L} + 1)+\mathcal{N}\log_{2}(\mathcal{N})(\mathcal{L}_{\bm{c}} + 1)$, homomorphic rotations have a complexity of $C(R)=\mathcal{O}(\mathcal{N}\log_{2}(\mathcal{N})\mathcal{L}_{\bm{c}}^2)$, and ciphertext-ciphertext multiplications have a complexity of $C(M)=4N(\mathcal{L}_{\bm{c}}+1) + C(R)$~\cite{poseidon}. 

The degree ($\mathfrak{p}$) of the polynomial approximation affects the depth of the circuit required for an activation function; computing $\varphi$ (or its derivative $\varphi'$) consumes $\ceil{\log_2(\mathfrak{p})}$ ($\ceil{\log_2(\mathfrak{p}-1)}$, respectively) levels. Thus, using a higher $\mathfrak{p}$ for polynomial approximation has two consequences to the computational cost: (i) it increases the depth of the circuit, hence the number of bootstraps, and (ii) results in more homomorphic multiplications. As discussed in Section~\ref{sec:packing}, the multi-dimensional matrix multiplication (Algorithm~\ref{algo:matrixmultiplicationCipher}) consumes 3 levels and yields an amortized cost of $\mathcal{O}(\delta)$ rotations per matrix (depending on $\delta$, see Appendix~\ref{sec:paramSelection}). The Example Configuration requires 6 matrix multiplications per timestep (8 for the output-stage timestep due to the multiplications with $\mathbf{V}$ and $\mathbf{V}^T$ which adds a constant factor to the complexity), thus yielding a computation complexity of $\mathcal{O}(6T\delta)$ rotations for a local iteration of $T$ timesteps and per ciphertext. Recall that \sys has the optimized execution option for the matrix multiplication as one can pre-compute the the linear transformations for the matrix multiplications \textit{per iteration} to re-use them for all timesteps. For the Example Configuration, this reduces the cost to $\mathcal{O}(6\delta)$ per local iteration, at a memory cost of $\times\mathbb{d}$ per matrix.
\section{Implementation Details} \label{sec:Implementation}
We implement \sys in Go~\cite{Go} and employ the Lattigo~\cite{lattigo} lattice-based (M)HE library. We rely on Onet~\cite{onet} to build a decentralized system 
and Mininet~\cite{mininet} to emulate a virtual network. Our experiments are performed on 10 Linux servers with Intel Xeon E5-2680 v3 CPUs running at 2.5GHz with 24 threads on 12 cores and 256 GB RAM emulating a virtual network with an average network delay of 20ms and 1Gbps bandwidth.

\section{Detailed Dataset Description}\label{sec:appendixDataset}
\descr{Hourly Energy Consumption (HEC)}: This dataset contains historical hourly energy consumption (in MegaWatts) from 12 major electricity distribution companies across the United States. The data was collected by PJM Interconnection LLC's website and is publicly available on Kaggle~\cite{hecKaggle}. The dataset contains 12 files, each corresponding to a distribution company and to a specific time period. Table~\ref{table:hec_dataset} shows further details about the names of these 12 companies and the number of samples in each file. The total number of samples is $1,090,167$. Note that for our experiments with $N{=}10$ data holders, we use the first 10 companies for the training, but we include samples from all companies in the test set. To capture the seasonality in energy consumption (i.e., trends with respect to daytime vs. nighttime, weekdays vs. weekends, winters vs. summers etc.), we incorporate on top of the energy consumption values 5 more features in the input space: hour, day-of-the-week, day-of-the-month, month and year ($d{=}6$). We also apply Min-Max scaling on the data to minimize the risk of explosion on the approximated activation functions. Additionally, for every file, we separately perform Min-Max scaling on the energy consumption values. This allows for a realistic distribution of the data among the multiple data holders (each data holder has one file, and does not know the range of the energy consumption values of other data holders). Finally, we process the dataset with a sequence of length $T{=}[10,20]$.

\begin{table}[t]
\small
\centering
\setlength{\tabcolsep}{1.8pt}
\begin{tabular}{lll}
\toprule
\multicolumn{1}{c}{Company Name} & \multicolumn{1}{c}{File Name~\cite{hecKaggle}} & \multicolumn{1}{c}{\#Samples} \\ 
\midrule
American Electric Power & {[}AEP\_hourly.csv{]} & 121,273 \\
Commonwealth Edison & {[}COMED\_hourly.csv{]} & 66,497 \\
The Dayton Power and Light C. & {[}DAYTON\_hourly.csv{]} & 121,275 \\
Duke Energy Ohio/Kentucky & {[}DEOK\_hourly.csv{]} & 57,739 \\
Dominion Virginia Power & {[}DOM\_hourly.csv{]} & 116,189 \\
Duquesne Light Co. & {[}DUQ\_hourly.csv{]} & 119,068 \\
East Kentucky Power Cooperative & {[}EKPC\_hourly.csv{]} & 45,334 \\
FirstEnergy & {[}FE\_hourly.csv{]} & 62,874 \\
Northern Illinois Hub & {[}NI\_hourly.csv{]} & 58,450 \\
PJM East Region & {[}PJME\_hourly.csv{]} & 145,366\\
PJM West Region & {[}PJMW\_hourly.csv{]} & 143,206\\
PJM Load & {[}PJM\_Load\_hourly.csv{]} & 32,896\\
\bottomrule
\end{tabular}
\captionsetup{width=\linewidth}
\caption{Detailed description of the HEC dataset.}
\label{table:hec_dataset}
\end{table}

\descr{Stock Prices (Stock):} This dataset contains historical daily stock prices and statistics for 10 companies, and was collected by the Yahoo! Finance website~\cite{stocks}. Similar to the HEC dataset, we perform Min-Max scaling on each file separately. Table~\ref{table:stocks_dataset} shows the company name for each file, the data collection period, and the number of samples per file. The total number of samples is 39,873. To prepare the dataset for the forecasting task, we pre-process and slice it using a sliding window of size $T{=}5$.\\

\descr{Inflation dataset (Inflation):} This dataset contains quarterly inflation rate based on consumer price index from 1998-Q1 to 2022-Q1 with 97 timesteps for 40 different countries and was collected from the Organisation for Economic Cooperation and Development~\cite{inflation}. We split the data into 40 different files for a realistic distribution. Each data holder receives 4 files (imbalanced setting) or the aggregate data is distributed evenly to the data holders (even setting). We perform Min-Max scaling on each file separately. Each file is processed with a sequence of length 8.

\descr{Breast Cancer Wisconsin (BCW):} This dataset~\cite{bcw} contains benign and malignant breast cancer samples. As the original dataset is centralized with 699 samples (576 training and 123 test samples), we randomly distribute the data among $N{=}10$ data holders each having 57-58 training samples for the even setting. For the imbalanced setting, one party has half of the training data and we evenly distribute the other half to the remaining parties. We perform Min-Max scaling on each file separately. BCW has 9 features that we treat as timesteps, i.e., $T{=}9$ (we also experimentally evaluated $T{=}1$ and $d{=}9$ but did not observe a significant difference in classification accuracy).

\begin{table}[t]
\small
\centering
\setlength{\tabcolsep}{1.6pt}
\begin{tabular}{llllc}
\toprule
\multicolumn{1}{c}{Company Name} & \multicolumn{1}{c}{File Name~\cite{stocks}} & \multicolumn{1}{c}{From} & \multicolumn{1}{c}{To} &  \#Samples \\
\midrule
Apple Inc. & {[}AAPL.csv{]} & 1980-12-12 & 2017-08-11 & 9,247 \\
Amazon.com, Inc. & {[}AMZN.csv{]} & 1997-05-15 & 2022-05-19 & 6,296 \\
Alibaba G.H.L. & {[}BABA.csv{]} & 2014-09-19 & 2022-05-19 & 1,931 \\
Facebook & {[}FB.csv{]} & 2012-05-18 & 2022-05-19 & 2,518 \\
Alphabet Inc. & {[}GOOG.csv{]} & 2004-08-19 & 2022-05-19 & 4,470 \\
Alphabet Inc. & {[}GOOGL.csv{]} & 2004-08-19 & 2022-05-19 & 4,470 \\
Netflix Inc. & {[}NFLX.csv{]} & 2002-05-23 & 2022-05-19 & 5,034 \\
Tesla, Inc. & {[}TSLA.csv{]} & 2010-06-29 & 2022-05-19 & 2,995 \\
Twitter, Inc. & {[}TWTR.csv{]} & 2013-11-07 & 2022-05-19 & 2,148 \\
Uber Tech., Inc. & {[}UBER.csv{]} & 2019-05-10 & 2022-05-19 & 764\\
\bottomrule
\end{tabular}
\captionsetup{width=\linewidth}
\caption{Detailed description of the Stock dataset.}
\label{table:stocks_dataset}
\end{table}

\section{Memory Complexity}\label{sec:memory}
Table~\ref{table:matrixMultScaling} displays the memory complexity of our multi-dimensional packing, as well as Jiang et al.'s~\cite{Jiang2018} and Sav et al.'s~\cite{poseidon} packing approaches.

\begin{table}[h]
\centering
\small
\setlength{\tabcolsep}{1.3pt}
\begin{tabular}{cccc}
\toprule
  &  ~\cite{Jiang2018} & ~\cite{poseidon} & Multi-dimensional \\ \midrule
$\mathcal{N}$  & $\mathcal{O}(n^{2})$ & $\mathcal{O}(n^{2})$ & $\mathcal{O}(\delta^{2})$\\ 
\#Ciphertexts  & $\mathcal{O}(1)$ & $\mathcal{O}(1)$ & $\mathcal{O}(n^{2}/\delta^{2})$ \\ 
Evaluation Keys (Size)  & $\mathcal{O}(n^{3})$ & $\mathcal{O}(n^{2}\log{n})$ & $\mathcal{O}(\delta^{3})$\\ \bottomrule
\end{tabular}
\captionsetup{width=\linewidth}
\caption{Comparison of the memory complexity between our multi-dimensional packing to Jiang et al.'s~\cite{Jiang2018} and Sav et al.'s (\poseidon)~\cite{poseidon} approaches, with respect to the matrix and sub-matrix dimensions $n$ and $\delta$.}
\label{table:matrixMultScaling}
\end{table}

\end{document}